# Dynamic Switching of Ferrocene and Plasmonic Interactions in Au/Self-Assembled Monolayer/Single Ag Nanocube Molecular Junctions.

Mariem BEN YOUSSEF,[1] Peeranuch POUNGSRIPONG,[2] Hugo BIDOTTI,[3] Halidou ABDOUL-YASSET,[4] Didier GIGMES,[3] Olivier MARGEAT,[2] Beniamino SCIACCA,[2] Judikael Le ROUZO,[4] David DUCHE[4] and Dominique VUILLAUME.[1]*

1) Institute for Electronics Microelectronics and Nanotechnology (IEMN), CNRS, Av. Poincaré, Villeneuve d'Ascq, France.
2) Centre Interdisciplinaire de Nanoscience de Marseille (CINaM), CNRS, Aix-Marseille University, Marseille, France.
3) Aix-Marseille University, CNRS, Institut de Chimie Radicalaire (ICR), Marseille, France.
4) Institut Matériaux Microélectronique Nanosciences de Provence (IM2NP), CNRS, Aix-Marseille University, Marseille, France.

\* Corresponding authors: dominique.vuillaume@iemn.fr

**Abstract.**

We report the redox switching of ferrocene moieties embedded in a double tunnel barrier plasmonic cavity. A self-assembled monolayer of ferrocenyl-alkylthiol is formed by click chemistry on an ultra-flat gold surface and a poly(vinylpyrrolidone) capped silver single nanocube, AgNC, is deposited. The AgNC is contacted by the tip of a conductive-AFM to study the electron transport properties in the dark and under light irradiation at the plasmonic resonance wavelengths. We observe a dual behavior in the current-voltage (I-V) characteristics in the dark: a large hysteresis loop at positive voltages and an hysteretic negative differential conductance (NDC) at negative voltages, due to the redox switching of ferrocene between its oxidized ($Fc^+$) and neutral ($Fc^0$) states. The I-V curves are analyzed by a combined Marcus-Landauer model. We determine the highest occupied molecular orbital of the $Fc^+$ and $Fc^0$ states at 0.54 and 0.42 eV below the Fermi energy, respectively, with a weak reorganization energy < 0.1 eV upon switching. Under plasmonic excitation, the hysteresis and NDC behaviors are no longer observed and the I-V characteristics of the Au-ferrocenyl-alkylthiol/AgNC junctions become similar to Au-ferrocenyl-alkylthiol SAMs. A virtual molecular orbital due to the plasmon-induced coupling (fast electron transfer) between the two redox states of the Fc is determined at 0.46 eV. This dynamic behavior opens perspectives in artificial synaptic devices for neuromorphic computing with the additional function to turn on/off this synaptic behavior on-demand by light.

**Keywords.** redox molecule, molecular switch, plasmonic, electron transport, molecular electronic

# INTRODUCTION

Molecular plasmonic devices belong to a field of research at the interface between physics, nanotechnology, materials science and molecular engineering. Plasmonic nanostructures are used to improve the light-harvesting efficiency in systems such as of solar cells,[1-3] or to improve the efficiency of light harvesting by nano-rectennas.[4-7] In data storage and communication technologies, plasmonic waveguides and circuits offer the potential of miniaturized components operating at optical frequencies.[1] In the field of molecular electronics, by incorporating molecules with specific chemical functions into plasmonic nanostructures, we can create devices that respond selectively to different stimuli with foreseen applications for sensors, photonics and optoelectronics.[8] Plasmon-molecule interactions are subtle resulting in a large variety of behaviors. Experimental and theoretical results are reviewed in Refs. 8-10. Many works have reported the plasmon-assisted increase of the electrical conductance in various molecular junctions (MJs). Among other results, plasmonic enhancement of the current[11] was reported for various molecules and device architectures: in 2D arrays of nanoparticles linked by alkylthiols,[12] oligo(phenylene vinylene)dithiols[13] and dithiol-Zn-porphyrins;[14] in "suspended-wire" molecular junctions connecting self-assembled monolayers of alkylthiols;[15,16] down to single-molecule junctions with diaminofluorene.[17] Most of the time, this conductance enhancement is explained by plasmon-assisted tunneling[18] but bolometric photoconductance can also come into play.[12] Conversely, the conductance of the molecular junctions is able to control quantum charge transfer plasmons.[19,20] The interaction of plasmons with molecular switches can also increase the switching speed (*e.g.*, a ≈ 50% faster switching for diarylethenes due to the plasmonic electric field enhancement),[21] as well as plasmon-induced *trans-cis* isomerization of azobenzene derivatives at visible light with a faster isomerization kinetics than under the usual UV light illumination (by at least a factor ≈10).[22] However, if the molecular spacer between the photochromic moiety and the nanoparticle (*e.g.*, typically an alkyl chain or a π-conjugated oligomer) is too short (< 4 carbon atoms in alkyl chains) the photoisomerization is quenched by the surface plasmon resonance.[23,24] Under plasmonic excitation, a redox organic nanowire (PEDOT, poly(3,4-ethylenedioxythiophene)) can be switched from its high to low conductance states with an on/off ratio of ≈ $10^3$ due to plasmon-induced hot-electron injection from the nanoparticle into the PEDOT nanowire that turns the PEDOT into its insulating reduced state.[25] In the case of a strong plasmon-molecule coupling and the creation of hybrid light-matter states,[26, 27] chemical and photochemical reactions are strongly modified slowing down the rate, or even completely suppressing, the photochemical reactions.[28, 29]

Ferrocene (Fc) is an archetype of redox molecules with potential applications in molecular electronic nanodevices (see recent reviews in 30, 31). The Fc molecule is an organometallic molecule featuring a single iron atom sandwiched between two cyclopentadienyl rings. The reversible Fe(II)/Fe(III) redox couple provides a very reliable and well-defined electrochemical property, the Fc molecule is stable under ambient conditions. The functionalization of the cyclopentadienyl rings is easy to tune the electrochemical properties of Fc and to allow its insertion in more complex molecular architectures. Focusing on molecular electronics, ferrocene molecules and derivatives have been used in rectifying molecular diodes,[32, 33] with a high rectification current ratio (ratio between the forward and reverse currents) up to $6.3 \times 10^5$ for a specific Fc-C≡C-Fc moiety.[34] The Fc-based molecular diodes were operated at microwave frequencies (≈20 GHz) with an extrapolated bandwidth of ≈ 600 GHz,[35] making them prone for high-frequency molecular electronics. As a single molecule junction, high conductances, near the conductance quantum $G_0$ ($7.75 \times 10^{-5}$ S), were reported (≈ 0.7-0.8 $G_0$) due to low-lying (i.e., very close to the Fermi energy of electrodes, *i.e.*, ≈ 30 meV) molecular orbitals of the Fc molecules coupled with the metal electrodes.[36, 37] A Fc-based light switchable single-molecule device was also recently demonstrated.[38] Upon a photooxidation reaction a direct metal-metal contact between the oxidized Fe(III) and the gold tip apex is formed (STM break junction in solution) leading to an enhanced conductance with respect of the Au tip/cyclopentadienyl contact (by a factor ≈ 15). Memory and switching devices were also investigated owing to the bistable redox states of Fc derivatives. Almost all the studied Fc-based molecular junctions used Au substrate as the bottom electrode and thiol chemistry to graft the molecules on the substrate. In the perspective to integrate Fc-based devices with the mainstream semiconductor technologies, Fc derivatives were also chemically attached on silicon substrates[39-41] for memory and light-stimulated switching applications.[42-44] Memories based on self-assembled monolayers (SAM) of Fc derivatives were reported with the advantage of operating at a low voltage (< 1 V) and a high density of charge storage (≈ 40 µC/cm$^2$) owing to a high surface coverage of Fc molecules (≈ $2.5 \times 10^{14}$ cm$^{-2}$) for densely packed SAMs.[45-47] More recently, Fc molecules were combined with 2D materials for exploring new concepts and nanoscale devices. Using a graphene monolayer as the top electrode, it was possible to decouple the redox electrochemistry and the charge transport in the molecular junction on either side of the graphene layer, resulting in a switching on/off ratio of ≈120, the neutral state having a higher conductance than the oxidized state.[48] Ferrocene derivatives were also used to functionalize 2D transition metal dichalcogenide (*e.g.*, $MoS_2$) layers and tune the electron transport properties of these 2D materials and devices (*e.g.*, $MoS_2$ transistors) depending on the Fc redox state.[49-51] Finally, Fc-based self-assembled diodes, embedded in plasmonic nanostructures, were proposed as a suitable platform for rectenna applications,[52, 53] which forms the backdrop to the work reported in the following.

      Here, we studied the redox switching of ferrocene moieties embedded in a double tunnel barrier plasmonic cavity and we compared the electron transport properties of this molecular device in the dark and under light irradiation at the plasmonic resonance wavelengths. In earlier works, hysteresis and negative differential conductance (NDC) in the current-voltage characteristics have been observed in experiments at the nanoscale (STM or C-AFM) on Fc-based MJs. Some authors have attributed the NDC behavior to resonant charge tunneling through a molecular orbital, like in semiconductor resonant tunnel diodes,[54, 55] while other works correlated the hysteresis and NDC with the Fc redox switching.[56-59] However, the Fc redox switching interactions with light in a single nanoparticle plasmonic cavity has not been reported to date. The plasmonic MJ was fabricated on ultra flat gold surfaces covered with a self-assembled monolayer (SAM) of ferrocenyl-alkylthiol. Silver nanocubes (AgNCs) which are capped by an ultra-thin layer of poly(vinylpyrrolidone) (PVP), were deposited at a low concentration on the SAM surface so as to obtain single isolated AgNC. The metallic tip of a conductive-AFM (C-AFM) is used to electrically contact a single Au-ferrocenyl-alkylthiol/AgNC to study its electron transport properties. In the dark, we observed a dual behavior in the current-voltage (I-V) characteristics. Two remarkable features were observed: a large hysteresis loop at positive voltages and an hysteretic NDC at negative voltages. We explained this dual behavior within the framework of a generalized electron transport theory combining the Marcus and Landauer approaches.[60-69] With the help of control experiments on the same SAMs (without the AgNC) and SAMs without the ferrocene moieties, these behaviors are explained by the redox switching of ferrocene between its neutral ($Fc^0$) and oxidized ($Fc^+$) states and by the double tunnel barrier structure of the device (alkyl chains and PVP layer). This double barrier configuration induces a weak coupling of the Fc with the electrodes, which increases the lifetime of the $Fc^0$ state, making it observable. Under

plasmonic excitation, the hysteresis and NDC behaviors are suppressed, and the current-voltage curves of the Au-ferrocenyl-alkylthiol/AgNC junctions are quite similar to those recorded on the same Au-ferrocenyl-alkylthiol SAMs directly contacted by the C-AFM tip near the AgNCs. We hypothesize that the plasmonic-enhanced electrical field in the cavity induces a coupling (fast electron transfer) between the two redox states of the Fc leading to a virtual molecular orbital in the Au-ferrocenyl-alkylthiol/AgNC. This dynamic behavior (hysteresis loop and NDC) open perspectives in artificial synaptic devices for neuromorphic computing,[70] with the additional function to turn on/off this synaptic behavior on-demand by light in our plasmonic Au-ferrocenyl-alkylthiol/single AgNC molecular junctions.

**EXPERIMENTAL METHODS.**

***Molecule synthesis and SAM junction fabrication.*** The molecules were commercially available or where synthesized as reported in Refs. 71, 72. The SAMs were made in two steps as reported elsewhere.[71, 72] We prepared ultra flat template-stripped gold surfaces ($^{TS}$Au), with rms roughness of ~ 0.3-04 nm, according to methods already reported.[73-76] The binary SAMs of octane-1-thiol and 11-azidoundecane-1-thiol were prepared on these $^{TS}$Au substrates in solution (1 mM of the two molecules at a 1:1 ratio in ethanol for ca. 1 day). The octane-1-thiol is used as a diluent to favor: i) a good organization of the alkyl chains in the SAMs and ii) an efficient grafting of Fc in the second step.[71, 72] The second step consists to attach the ferrocene moieties by click chemistry on the $N_3$ terminal groups of the 11-azidoundecane-1-thiol available at the SAM surface (see details in Refs. 71, 72). The detailed structural characterizations of these SAMs were reported elsewhere (cyclic voltammetry, IR spectroscopy, XPS, ellipsometry, contact angles measurements, refer to "SAM3" and "SAM4" in Refs 71,72 for the binary SAMs of octane-1-thiol and 11-azidoundecane-1-thiol and with the ferrocene moieties after the click reaction, respectively). The thicknesses of the SAMs (spectroscopic ellipsometry measurements see details in the Supporting Information) are 1.7 ± 0.2 nm (without the Fc) and 2.8 ± 0.2 nm with the Fc in reasonable agreement with the expected length of the molecules (1.7 nm for $SC_{11}N_3$, 2.6 nm when the ferrocene moiety is added, geometry optimized MM2) and previously reported values for the same SAMs.[71, 72] The topographic AFM images (Fig. S1) of the SAMs (with and without Fc) reveal flat and homogeneous SAMs, free of gross defects (neither pinhole nor aggregate). The measured rms roughness value for the surface of the $^{TS}$Au-S-$C_{8/11}$ SAM is 0.39 nm, and 0.66 nm for the $^{TS}$Au-S-$C_{8/11}$-Fc SAM (see the Supporting Information). Additional XPS measurements were conducted in this work to better characterize the grafting of the molecules on the Au substrate and to determine the charge states of the ferrocene moieties. They clearly show (see the Supporting Information) the characteristic feature of the S-Au bond and from the Fe2p region, we estimated a ratio [Fe(III)]/[Fe(II)] ≈ 6.

***Silver nanocube synthesis and deposition on SAMs.*** Silver nanocubes were synthesized using polyol synthesis.[77] First, 25 mL of ethylene glycol was added to a round-bottom flask and preheated to 160 °C with stirring at 300 rpm. After reaching 160 °C, 300 µL of 3 mM sodium hydrosulfide was added. Two minutes later, 2.5 mL of 3 mM hydrochloric acid and 6.25 mL of 20 mg/mL polyvinylpyrrolidone (PVP) were added. Another 2 minutes later, 2 mL of 282 mM silver trifluoroacetate was introduced. The reaction was quenched in an ice-water bath after 1.5 hours for 50 nm nanocubes and 2 hours for 60 nm nanocubes. The size of the nanocubes was determined using UV-Vis spectrometry. The nanocubes were collected by centrifugation and washed with acetone and deionized water. They were re-dispersed in water. For further processing, 10 µL of the nanocube stock solution was mixed with 100 µL of 100 mM sodium borohydride in deionized water for 3-5 minutes to remove residual PVP. Then, 2-5 µL of this mixture was drop-cast on Au/SAMs substrates and pressed with a glass coverslip for 2 minutes. The samples were rinsed with DI water, dried with $N_2$ gas, resulting in sparse single nanocube deposition.

***Conductive-AFM measurements.*** For the C-AFM measurements, we used a Dimension Icon microscope (Bruker) installed in an air-conditioned laboratory (T*amb* = 22.5 °C, a relative humidity of 35-40%). We used a conductive PtIr metal-plated tip (model RMN-12PT400B from Bruker). The voltage was applied on the $^{TS}$Au substrate, the C-AFM tip grounded. To contact an individual AgNC, we followed a procedure already reported to electrically characterize NCs (of about the same size but of a different chemical nature).[78] In brief, a topographic image in tapping mode (with the PtIr tip) is used to target an isolated AgNC, then we switched from the tapping to contact mode with the tip almost at a stationary point above a given AgNC, the x- and y-scans turned off (see details of the method in Ref. 78). We used a loading force of ca. 11-15 nm to insure a good mechanical contact on the AgNC through the ultra-thin PVP layer. 20 I-V traces were acquired, repeated on 5 different AgNC to build the raw data set (100 I-V traces). For the I-V measurements directly on the SAMs, we defined a square grid of 10 × 10 points (pitch of 50 to 100 nm and at each point, a back and forth I-V curve was acquired (given 200 traces in the raw data set). A smaller loading force is used in that case (3-5 nN). All the I-Vs were acquired at a voltage sweep rate of ≈ 0.5 V/s (otherwise specified). Then, the raw set of I-V data was inspected and some I-V curves were discarded from the analysis (see details in the Supporting Information). The final number of I-V traces retained for analysis is indicated in the figure captions.

*Light illumination.* We used LEDs (from ThorLabs, LED MBB1F1 broadband 470-850 nm and LED M470F4 at 470 nm with a FWHM of 20 nm) coupled with a multimode optical fiber (BF20HSMA01, ThorLabs). The output of the optical fiber is placed close to the C-AFM tip (a few cm) and the light has an angle of incidence of ≈ 45°. In this geometrical configuration, the incident light intensity is ≈ 28 W/m$^2$ for both the broadband LED (almost constant in the 500 - 800 nm range) and for the M470F4 LED measured at 470 nm (measured with a ThorLabs power meter PM204, the silicon S120VC photodetector being at the position of the sample).

## RESULTS.

The MJ is a binary SAM of octane-1-thiols and 11-azidoundecane-1-thiols chemisorbed on template-stripped Au ($^{TS}$Au), functionalized in a second step with ferrocene moieties by click chemistry as reported in Ref. [72] (See Experimental Methods), referred to as $^{TS}$Au-S-C$_{8/11}$-Fc. AgNCs (60 nm side length, capped with a 1-3 nm thick layer of PVP)[79] were synthesized (see Experimental Methods), deposited from the solution on the SAMs and individually connected by a conductive-AFM tip (Fig. 1a, see Experimental Methods). The PVP layer provides a coating barrier preventing the easy oxidation of AgNC. The octane-1-thiol is used as a spacer to control the density of Fc moieties in the SAM,[72] and to minimize the π-π interactions between neighboring Fc, which disturb the energetics of the Fc in the molecular junctions.[80] Figures 1b-c show a topographic AFM image of few individual AgNCs deposited on the $^{TS}$Au-S-C$_{8/11}$-Fc SAM and the corresponding height profile, clearly indicating the presence of individual AgNCs. Ellipsometry and topographic AFM measurements show that homogeneous and compact SAMs are formed on the $^{TS}$Au substrates (see Experimental Methods and the Supporting Information).

Figure 2a shows the back-and-forth current-voltage (I-V) characteristics recorded on 5 $^{TS}$Au-S-C$_{8/11}$-Fc/AgNC/C-AFM tip junctions (see details in Experimental Methods). The red traces are the forward traces (-1 V → 1 V), the backward traces (1 V → -1 V) are in blue. At positive voltages, we observed a large hysteresis between the back-and-forth I-V traces and a NDC peak for the negative voltages (only for backward traces, *i.e.*, an hysteretic NDC). The mean on/off ratio (*i.e.* the current ratio of the mean backward trace over the mean forward one) is high with a maximum value of ≈ 700 (at 0.7 V, see Fig. S6 in the Supporting Information). The mean ratio between the peak current (at ≈ -0.1 V) and the valley current (at ≈ -0.4 V) is ≈ 3 (Fig. 2a). The NDC peak is observed only after having applied positive voltages on the junctions (*i.e.*, no NDC for a back and forth voltage scan limited between -1 V and 0 V, Fig. S7 in the Supporting Information). As a control experiment, the same SAMs were measured without AgNC by gently contacting the C-AFM tip directly on the SAM near an AgNC (details in Experimental Methods). Figure 2b shows the I-V traces acquired in the same conditions as for the $^{TS}$Au-S-C$_{8/11}$-Fc/AgNC/C-AFM tip junction (back and forth from -1 to 1 V, at the same voltage scan rate ≈ 0.5 V/s). The large hysteresis and the NDC feature are no longer observed in that case. To check the role of the ferrocene moiety, we also measured the I-Vs of $^{TS}$Au-S-C$_{8/11}$ SAMs directly contacted by the C-AFM tip and with the AgNCs (Fig. 4). Again, no large hysteresis, nor NDC peak, are observed.

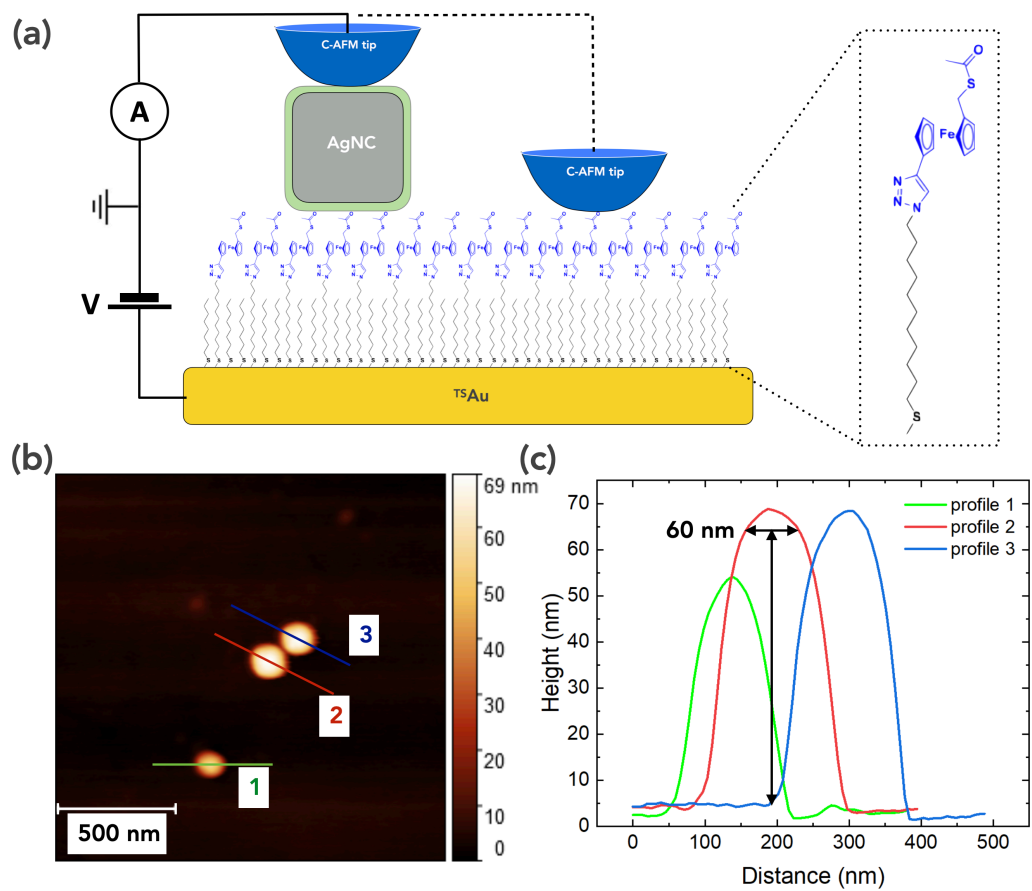

*Figure 1. (a)* Scheme (not on scale) of the $^{TS}$Au-S-C$_{8/11}$-Fc SAM connected by AgNC/C-AFM tip (left) or directly by the C-AFM tip (right), and of the alkyl-Fc molecule. PVP around the AgNC is symbolized in light green. *(b)* Topographic images of 3 individual AgNC (nominal size of 60 nm) deposited on the $^{TS}$Au-S-C$_{8/11}$-Fc SAM. *(c)* Height profiles (not deconvoluted from the tip shape). The black arrows indicate the nominal size. Two AgNCs have a height and diameter of ca. 60 nm.

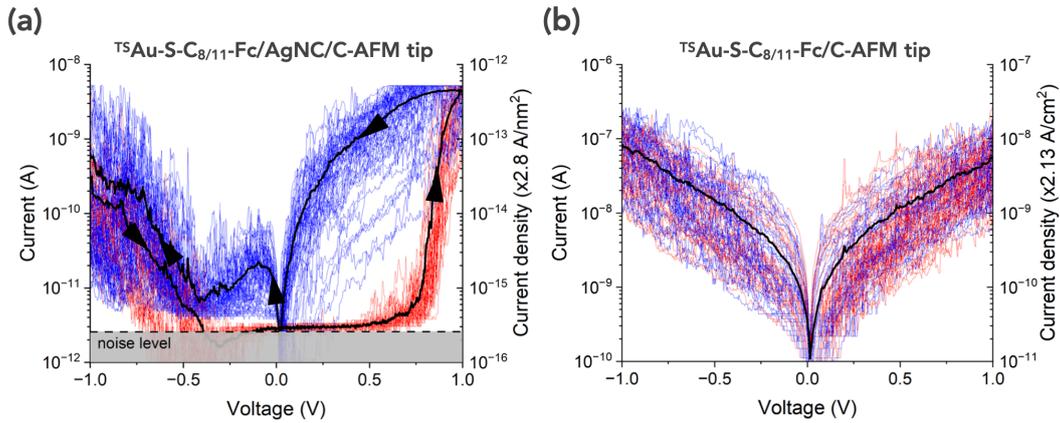

***Figure 2. (a)*** *I-V data set (90 traces) of the $^{TS}$Au-S-C$_{8/11}$-Fc/AgNC/C-AFM tip junctions. The dark lines are the mean Ī-V for the forward (red) and backward (blue) traces.* ***(b)*** *I-V data set (156 traces) of the $^{TS}$Au-S-C$_{8/11}$-Fc/C-AFM tip junction. The back line is the mean Ī-V from all the traces. The current density scales are calculated with the estimated areas of the AgNC and C-AFM tip, respectively (see the Supporting Information).*

The I-V data sets were analyzed by a single energy level (SEL) model.[81, 82] This model assumes that the electron transport hough the molecular junction is mediated by the molecular orbital, HOMO or LUMO depending on the molecule that is the closest to the Fermi energy of the electrodes. An analytical equation (see the Supporting Information, Eq. S4) was fit on all the I-V traces of the data set to obtain histograms of the energy level $\varepsilon_0$ (with respect to the Fermi energy), see details on the fit protocol in the Supporting Information. Figure 3 shows the energy level $\varepsilon_0$ histograms for the $^{TS}$Au-S-C$_{8/11}$-Fc/AgNC/C-AFM tip junctions and $^{TS}$Au-S-C$_{8/11}$-Fc/C-AFM tip junctions obtained from the I-V data sets given in Fig. 2. Note that for the $^{TS}$Au-S-C$_{8/11}$-Fc/AgNC/C-AFM tip junctions, only the backward traces from 1 V to 0 V were used (see details in the Supporting Information, the forward traces cannot be used to the fit due to the large plateau at the sensitivity limit of the experimental setup, the backward curves from 0 to -1 V, which show the NDC effect, cannot be fit with this model and another model was used in that case, *vide infra* in the discussion section). Under this limit, the ET through the $^{TS}$Au-S-C$_{8/11}$-Fc/AgNC/C-AFM tip junction is characterized by a lower energy of the MO $\varepsilon_0$ (0.42 ± 0.05 eV) than the same SAM directly contacted by the C-AFM tip ($\varepsilon_{0-1}$ = 0.54 ± 0.07 eV). Thus, the top contact with the AgNCs clearly modify the ET of the $^{TS}$Au-S-C$_{8/11}$-Fc SAM.

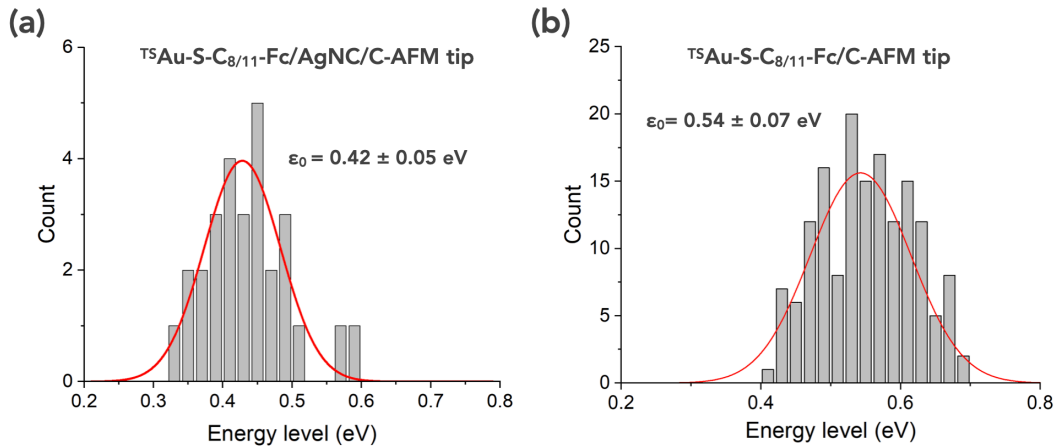

*Figure 3.* Histograms of energy level $\varepsilon_0$ of the molecular orbitals involved in the electron transport. The lines are the fit with Gaussian distributions with the values (mean ± standard deviation) given in the panels: *(a)* $^{TS}$Au-S-C$_{8/11}$-Fc/AgNC/C-AFM tip junctions, from I-V datasets shown in Fig. 2a (fit only on the backward 1 → 0 V traces, see the Supporting Information), *(b)* $^{TS}$Au-S-C$_{8/11}$-Fc/C-AFM tip junctions, from I-V datasets (full voltage range) shown in Fig. 2b.

To further analyze the role of the AgNCs, we have also done the same experiments with $^{TS}$Au-S-C$_{8/11}$ SAMs without the click chemistry step. Figure 4 shows the I-V data sets and the energy level $\varepsilon_0$ histograms for the $^{TS}$Au-S-C$_{8/11}$/AgNC/C-AFM tip and $^{TS}$Au-S-C$_{8/11}$/C-AFM tip junctions (see details for SEL model fits in the Supporting Information). In both cases, almost the same I-Vs and the same energetics of the molecular junctions were obtained, with $\varepsilon_0$ = 0.71 ± 0.09 eV and $\varepsilon_0$ = 0.76 ± 0.08 eV with and without the AgNCs, respectively. We only note a small hysteresis with the AgNCs at positive voltages (Fig. 4a), which likely induces the larger distribution in the histograms with a tail at high $\varepsilon_0$ values (Fig. 4c).

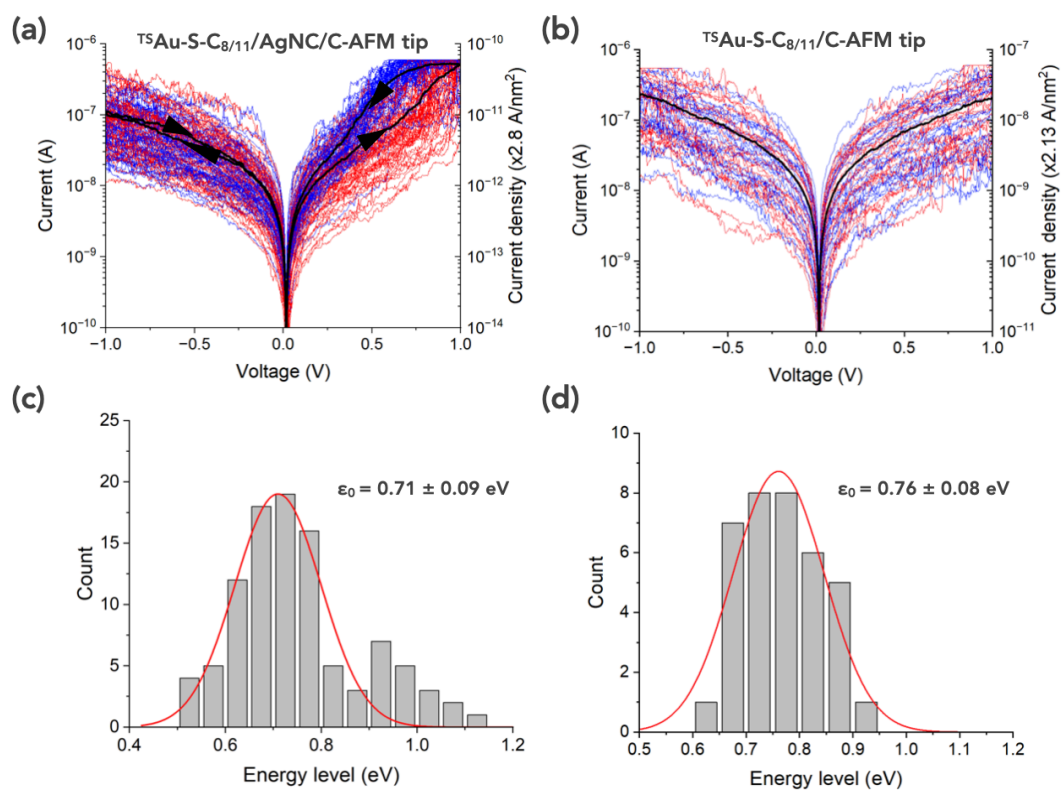

*Figure 4. (a)* I-V data set (120 traces) of the $^{TS}$Au-S-C$_{8/11}$/AgNC/C-AFM tip junction. The dark lines are the mean Ī-V for the forward (red) and backward (blue) traces. *(b)* I-V data set (49 traces) of the $^{TS}$Au-S-C$_{8/11}$/C-AFM tip junction. The back line is the mean Ī-V from all the traces. The current density scales are calculated with the estimated areas of the AgNC and C-AFM tip, respectively (see the Supporting Information). *(c-d)* Histograms of energy level $\varepsilon_0$ of the molecular orbitals involved in the electron transport. The lines are the fit with Gaussian distributions with the values (mean ± standard deviation) given in the panels: *(c)* $^{TS}$Au-S-C$_{8/11}$/AgNC/C-AFM tip junctions, from I-V datasets shown in Fig. 4a, *(d)* $^{TS}$Au-S-C$_{8/11}$/C-AFM tip junctions, from I-V datasets shown in Fig. 4b.

Under light illumination (see Experimental Methods) of the $^{TS}$Au-S-C$_{8/11}$-Fc/AgNC/C-AFM tip junctions, the I-V hysteresis and NDC behavior are suppressed (Fig. 5). The I-V dataset measured under light is similar as the one measured for the same SAMs without the NC, and the SEL analysis gives an energy distribution of the molecular orbital at $\varepsilon_0$ = 0.46 ± 0.05 eV (Fig. 5c). We have used a broadband light source (wavelengths 450-870 nm, see Experimental Methods) because the plasmonic resonances for these structures are not accurately known. The exact value of the plasmonic resonance wavelengths may vary from single device to device, since the SAM thickness may slightly vary from place to place as well as the PVP thickness[79] around the AgNC (1.5 ± 0.9 nm). Nevertheless, the plasmonic excitations are expected to be in the range 600-800 nm[79, 83] (see the Supporting Information for simulations). When the same $^{TS}$Au-S-C$_{8/11}$-Fc/AgNC/C-AFM tip junctions were illuminated with a small bandwidth light source (470 nm, FWHM 20 nm, see Experimental Methods), *i.e.* outside the plasmon excitation range, both the hysteresis and NDC behaviors are unaffected (Fig. S10 in the Supporting Information). We conclude that the excitation of plasmon modes is required to observe the suppression of the hysteresis and NDC.

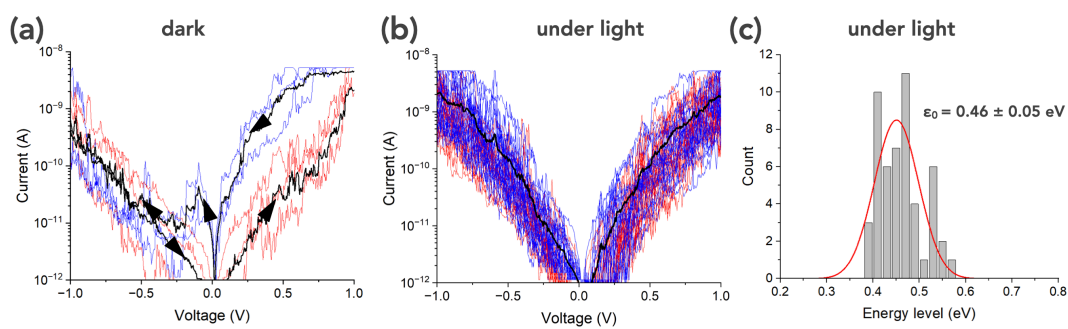

*Figure 5*. *(a)* Typical I-V curves in the dark of the $^{TS}$Au-S-C$_{8/11}$-Fc/AgNC/C-AFM tip junctions (other NCs than in Fig. 2). Only a few traces were recorded to check the hysteresis and NDC. The dark lines are the mean $\bar{I}$-V for the forward (red) and backward (blue) traces. *(b)* I-V data set (45 traces) of the $^{TS}$Au-S-C$_{8/11}$-Fc/AgNC/C-AFM tip junction under light illumination (470-850 nm wavelength). The back line is the mean $\bar{I}$-V from all the traces. *(c)* Histograms of energy level $\varepsilon_0$ of the molecular orbitals involved in the electron transport under illumination (SEL model applied on data shown in Fig. 5b). The lines are the fit with Gaussian distributions with the values (mean ± standard deviation) given in the panel.

We also note that the same broadband light illumination on the same SAMs (without the AgNCs) has no effect on the I-V behavior (Fig. S11). Under our conditions (wavelength and light intensity, see Experimental Methods), there is no detectable photocurrents in the $^{TS}$Au-S-C$_{8/11}$-Fc/C-AFM tip junctions: no HOMO-LUMO gap photo-excitation (the Fc HOMO-LUMO gap is too large ≈ 2.8 eV), nor measurable internal photoemission at the molecule/electrode interfaces. Thus, we assume that photocurrents though the SAM play no role and that the observed suppression of the hysteresis and NDC behaviors is related to the plasmonic structure. With an incident light intensity of ≈ 28 W/m$^2$ (see Experimental Methods), the free space ac electric field is 1.5x10$^{-7}$ V/nm (or a light-induced voltage of ca. 0.4 µV though the ca. 2.8 nm thick -S-C$_{8/11}$-Fc SAM) and no plasmon-assisted tunneling is expected under the C-AFM tip even with a typical plasmon amplification factor of the electric field of 10$^2$ - 10$^4$ in the Au surface/SAM/C-AFM tip MJs.[7, 16, 84, 85]

**DISCUSSION.**

We first briefly discuss the results for the $^{TS}$Au-S-C$_{8/11}$ SAMs. The molecular orbital at ≈ 0.71 - 0.76 eV (Fig. 4) is in agreement with several results for SAMs of alkylthiols on Au.[86-88] Whether this molecular orbital is the HOMO or the LUMO of the alkylthiol chains,[87, 89, 90] it is at an energy (with respect to the electrode Fermi energy) larger than the molecular orbital of the ferrocene (see below) and the electron transport through the alkylthiol spacers is off-resonant tunneling, which cannot explain the hysteresis and NDC behavior. The small hysteresis with the AgNCs at positive voltages (Fig. 4a) and the tail at high $\varepsilon_0$ values (Fig. 4c) is likely due to charging/impurity in the PVP tunnel barrier. For the $^{TS}$Au-S-C$_{8/11}$-Fc SAMs directly connected with the C-AFM tip, the molecular orbital at $\varepsilon_{0-1}$ = 0.53 ± 0.08 eV (Fig. 3b) is consistent with the energy position of the Fc HOMO as determined previously for the same SAMs by CV measurements.[72] It is also in agreement with several other reports for alkyl-Fc SAMs.[80, 91, 92] Further DFT simulations should confirm the energetics and electron transport properties of these MJs. A majority of the ferrocene are in their oxidized state Fc$^+$ in these SAMs (see XPS measurements in the Supporting Information, we estimated a ratio [Fc$^+$]/[Fc$^0$] ≈ 6) exposed to ambient conditions,[35, 80] thus we ascribe the level at 0.54 eV to the HOMO of the oxidized state Fc$^+$.

The hysteresis and NDC effects are due to the double tunnel barrier structure of the device (alkyl chains and PVP layer on each side of the Fc layer). Several works have already reported I-V hysteresis loops[56, 93-95] and NDC[54, 55, 57-59, 96-98] in MJs with redox molecules.[99] In the case of Fc-based MJs, these earlier experiments have attributed the NDC behavior to resonant charge tunneling through a molecular orbital, by analogy with semiconductor resonant tunnel diodes.[54, 55] Other works ascribed the hysteresis and NDC with Fc redox switching.[56, 58, 59] More generally, these redox-mediated electron transport behaviors can be understood in the framework of combined Marcus and Landauer theories that have been developed by several groups, *i.e.*, Kuznetsov, Ulstrup and coworkers,[60-63], Migliore and Nitzan,[64-66], Sowa, Marcus and colleagues.[67-69]

The hysteresis and NDC behaviors shown in Fig. 2a are explained considering 4 electron transport schemes (Fig. 6).

- Scheme I (Figs. 6a and c). Staring the measurement at -1V, and assuming that the Fc is initially in its oxidized state, $Fc^+$, with a HOMO at ≈ 0.54 eV (Fig. 6b) below the electrode Fermi energy (*vide supra*) the electron transport occurs coherently off resonance up to a threshold $V_{th}$ ≈ 0.7 V.

- Scheme II (Figs. 6a and d). Above this ≈ 0.7 V threshold, the HOMO approaches the energy window defined by the difference of the Fermi energies of the two electrodes and the electron transport takes place by sequential hopping processes (curved black arrows) and the $Fc^+$ moieties switch to a neutral state $Fc^0$ (symbolized by the upward white arrow in Fig. 6d), thus a second conduction channel is gradually opened (light red arrows) whereby the electron transport is coherently resonant. This situation has been theoretically described with a two conduction channels model where a "fast" channel (here $Fc^0$ with the closest HOMO with respect to the Fermi energy) mainly imposes the measured current, and a "slow" channel (here $Fc^+$ with a deeper HOMO) determines the charge state of the molecule.[66]

- Scheme III (Figs. 6a and e). The $Fc^0$ state has a HOMO energy closer to the electrode Fermi levels, ≈ 0.42 eV (as determined above, Figs. 3a, 6b), a larger current is measured through the molecular junction during the backward scan from 1 V to 0 V. The smaller energy of the $Fc^0$ HOMO (with respect to the Fermi energy) is opposite to the classical Gerischer model (oxidized state above the neutral one),[100, 101] but this model holds for a semiconductor or metal/redox molecule interface directly in contact with an electrolyte. This is not likely relevant here for a solid-state device with two insulating barriers between the redox molecule and the electrodes. Our energetic picture is in agreement with combined electrochemistry and UPS/XPS experiments on similar Fc SAM that show the same trend as proposed here.[102] We also note that a similar behavior was observed (UPS experiments and DFT calculations) for a sub-monolayer of $C_{60}$ molecules deposited on a thin insulating $MoO_3$ on Au.[103] The cationic $C_{60}$ molecules have a higher ionization energy than the neighboring neutral ones due to on-site and inter-site Coulomb interactions.

- Scheme IV (Figs. 6a and f). At very low negative bias, the $Fc^0$ and $Fc^+$ HOMOs are close to the Fermi energy and the two conductance channels are competing. Electron detrapping from the $Fc^0$ state switches the molecules to $Fc^+$ state while an electron trapping from the Au electrode switches back to the $Fc^0$ state until all the ferrocenes turn to the $Fc^+$ at larger negative voltages (V < -0.4V) and the system returns to scheme I.

Within the theoretical framework of a generalized Marcus-Landaueur electron transport theory, both I-V hysteresis loops and NDC peaks, qualitatively similar as the ones described in Fig. 2, have been simulated (*e.g.* see Figs. 5 and 9 in Ref. 66). However, these simulations required sophisticated computations.[66] Nevertheless, the $Fc^+/Fc^0$ switching (part II of the I-V curve in Fig. 6a) can be fit by the analytical expressions given by Eqs. (12, 33, 34) developed in Ref. 64. The fit is shown in Fig. 7a with an energy for the redox $Fc^+/Fc^0$ transition at 0.46 eV and a reorganization energy $\lambda \approx 0.02$ eV (see details in the Supporting Information). The weak reorganization energy suggests that only the inner-sphere reorganization is involved ($\approx 0.03$ eV according to Ref. 98) in this switching process.

For the 0 to -1 V scan, at very low bias (up to the NDC peak at $\approx$ - 0.1 V), the electron transport still occurs through the $Fc^0$ HOMO since we note a similar level of current as the 1 V to 0 V branch (light red arrows in Fig. 6f). Between $\approx$ - 0.1 V and $\approx$ -0.4 V (the valley of the NDC), we assume a competition between electron detrapping from the $Fc^0$ to the AgNC (switching back to the $Fc^+$ state) and electron trapping from the Au substrate (favoring $Fc^0$ state), scheme IV, symbolized by the double white arrow in Fig. 6f and the total current is fixed by the weighted contributions of two conductions channel depending on the concentrations of the two Fc charge states, which evolves with the applied voltage. This is consistent with two-level fluctuations noise measurements on alkyl-ferrocene molecular junctions whereby the rate of Fc oxidation and the rate of $Fc^+$ reduction are almost equal at a low voltage $|V| \approx 0.25$ V.[104] Increasing the voltage (abs value), the concentration of $Fc^0$ gradually decreases while the one of $Fc^+$ increases. This dynamic switching between the $Fc^0$ and $Fc^+$ states give rise to the observed NDC peak.

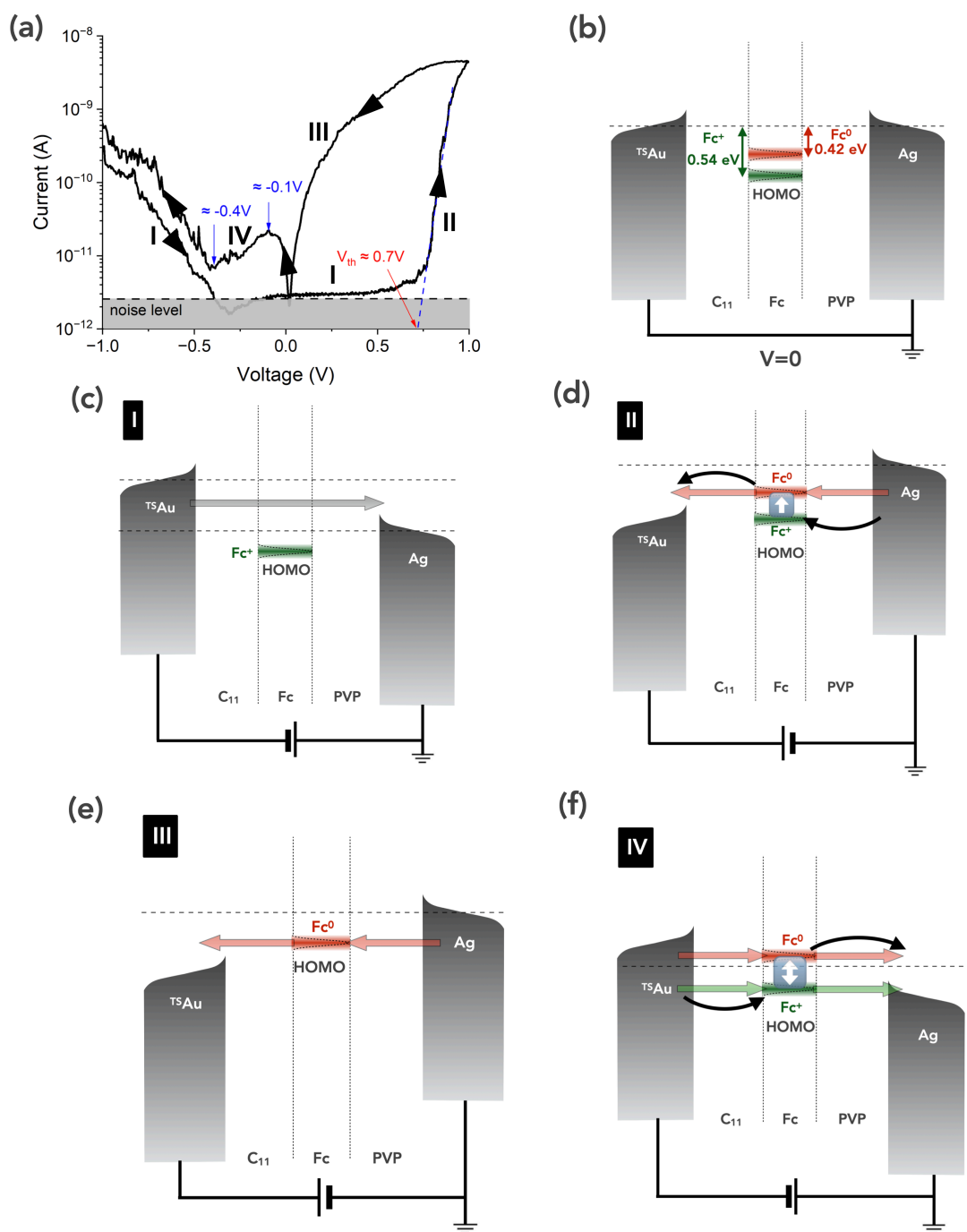

***Figure 6. (a)*** *Mean Ī-V of the $^{TS}$Au-S-C$_{8/11}$-Fc/AgNC/C-AFM tip junctions (from Fig. 2a) and the electron transport scenarios associated to the different parts of the Ī-V.* ***(b)*** *Schematic energy scheme of the $^{TS}$Au-S-C$_{8/11}$-Fc/AgNC junction at zero bias with the energy levels of the molecular orbitals as discussed in the text. The molecular orbitals of the alkylthiol C$_{11}$ and PVP layers are*

*omitted for clarity. Both have a large band gap (ca. 8-9 eV and 5-6 eV for the alkylthiol and PVP, respectively) and the electron transport through these spacer layers is dominated by off-resonant tunneling. **(c)** Electron transport scenario I. From - 1 V to $V_{th}$ there is no molecular orbital between the Fermi energy of the electrodes (or energy window). Note that in panels b and c, the $Fc^+$ energy level is stable below the Fermi energy since the molecule is weakly coupled to the electrodes (weak electron transfer) in this double tunnel barrier structure and the level is not resonant within the energy window (panel c). **(d)** Electron transport scenario II. Above $V_{th}$, the $Fc^+$ energy level enters the energy window defined by the applied voltage and starts switching to $Fc^0$ adding a resonant electron transport channel. **(e)** Electron transport scenario III. A majority of ferrocene moieties are in the $Fc^0$ state. The electron transport occurs through the HOMO of $Fc^0$, which is closer to the electrode Fermi energy than the one of $Fc^+$. **(f)** Electron transport scenario IV. At low negative bias (-0.4 to 0 V) the dynamic switching between $Fc^0$ and $Fc^+$ gives rise to the NDR peak (see text) until all the ferrocenes turn to the $Fc^+$ at larger negative voltages (V<-0.4V) and the system returns to scheme I.*

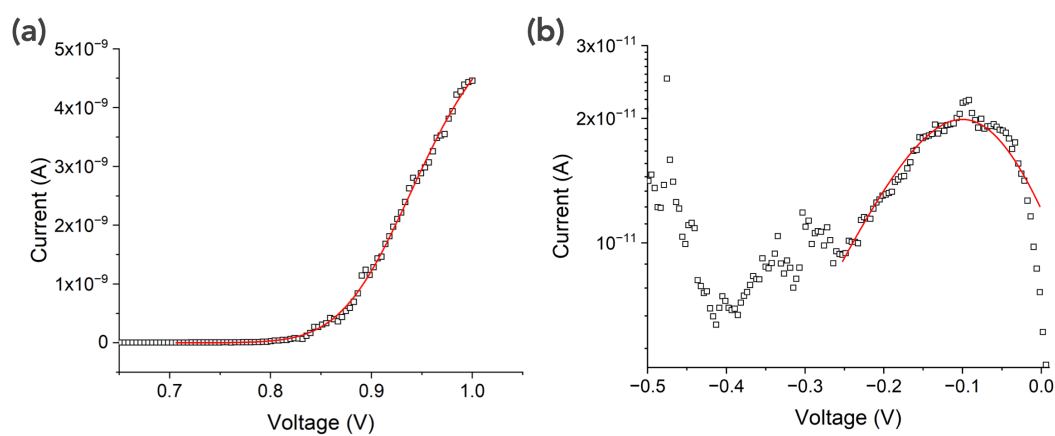

*Figure 7. **(a)** Fit of the Fc redox switching (part II of the I-V in Fig. 5a) and **(b)** fit of the NDC peak (part IV of the I-V in Fig. 5a) with analytical equations based on Marcus ET as developed in Ref. 64.*

We were able to fit the NDC peak (Fig. 7b) considering such a two-channel system and the equation 28 in Ref. 64 (see details in the Supporting Information). The energies for the two channels are $\varepsilon_1 \approx 0.2$ eV and $\varepsilon_2 \approx 0.04$ eV, respectively with a reorganization energy $\lambda \approx 0.1$ eV. These values are not exactly the same as deduced above from the $Fc^+/Fc^0$ switch (Fig. 7a), but the mechanisms and model are more complex for the NDC effect (see Supporting Information). Moreover, given the restricted voltage range used for the fit (from V = -0.25 to 0 V), these values can be considered as a rough estimate. However, we note that the "fast" second conduction channel is almost resonant with the Fermi energy of the electrodes, and that these values are roughly consistent with the ones (energy 0.17 eV, $\lambda \approx 0.3$ eV) derived from real-time redox switching detection measurements, for which a competition between the oxidation and reduction takes place as for the NDC.[104] Below $\approx$ -0.4 V, the ferrocene molecules have completely switched back to the $Fc^+$ state and the current returns to an almost similar curve as for the -1 V to 0 V scan. These effects are observed because the lifetime of the $Fc^0$ state is long enough (with respect to the voltage sweep time) due to the presence of the double tunnel barrier. The observation of these hysteretic NDC and large hysteresis loop is made possible because the MJs are in the weak electrode coupling regime, enabling long-lived redox states. When the C-AFM is directly contacting the Fc moiety, this $Fc^0$ lifetime is too short and we do not observe the hysteresis and NDC effects. An intermediate case was observed for samples of a second batch of $^{TS}Au$-S-$C_{8/11}$-Fc/ AgNC samples (fabricated with the same protocol). For these samples, the level of current is higher than for the samples shown in Fig. 2a, but the hysteresis at V > 0 has a weaker amplitude and the NDC effect at V < 0 is reduced (Fig. S9 in the Supporting Information, the backward mean $\bar{I}$-V just show a bump because only a small fraction ($\approx$ 10%) of the I-V traces display a marked NDC effect). Due to the large dispersion of the PVP layer capping the AgNC (transmission electron microscope measurements: 1.5 ± 0.9 nm)[79], the currents through the PVP layer span over large values (from $10^{-9}$ A to above $5\times10^{-7}$ A at +/-1 V, Fig. S8 in the Supporting Information). We assume that the higher level of current for batch #2 samples ($\approx 10^{-8}$ - $10^{-7}$ A at +/- 1 V) comes from AgNCs with a thinner PVP layer (varying between 1 and 3 nm, see Ref. 79) than for batch #1 samples (I $\approx$ $10^{-10}$ - $10^{-8}$ A at +/- 1 V). In that case, the Fc moiety is less isolated from the Ag electrode and the $Fc^+/Fc^0$ switching is less observable. The determination of the switching kinetics would require detailed voltage pulse experiments. Here we can only estimate that the switching occurs in less than $\approx$ 600 ms (i.e., the voltage window of $\approx$ 0.3 V for zones II and IV in the I-V, Fig. 6a, divided by the voltage sweep rate of 0.5 V/s). This upper limit is not in disagreement with the switching time constant of $\approx$ 100 ms determined from CV measurements.[72] The asymmetric behavior (hysteresis at V > 0 vs. NDC at V < 0), as well as the slightly different values for the redox and reorganization energies, are likely due to the asymmetry of the tunnel barriers (electron injection through PVP at V>0 and undecanethiol at V<0), the exact density of charges in the ferrocene layer (when in the $Fc^+$ state), resulting in different potential landscape across these barriers, all these parameters determining how the HOMOs of $Fc^0$ and $Fc^+$ states align with the electrode Fermi energy windows imposed by the applied voltage.

Under plasmon excitation, it was theoretically predicted that the interaction of a surface plasmon polariton and a two-level molecular system can reduce the conductance through the molecular junctions.[105, 106] Such a current decrease is not clearly observed (Fig. 5). We propose that the plasmon-enhanced field in the $^{TS}$Au-S-$C_{8/11}$-Fc/AgNC cavity (a factor *ca.* 70-140, see the Supporting Information) increases the $Fc^+$/$Fc^0$ redox switching rates (as observed for molecular photochromic switches)[21] and that the switching is no longer observable since the state lifetimes are faster than the typical time-scale of the I-V measurements as it is the case for the $^{TS}$Au-S-$C_{8/11}$-Fc/C-AFM tip junctions. The I-V dataset of the $^{TS}$Au-S-$C_{8/11}$-Fc/AgNC/C-AFM tip under light (Fig. 5b) was analyzed with the SEL model, leading to a molecular orbital energy distribution (Fig. 5c) $\varepsilon_{0,light}$ = 0.46 ± 0.05 eV. We note that this energy level lies between the $Fc^+$ (at 0.54 eV) and the $Fc^0$ (at 0.42 eV) states determined in the dark (Fig. 6b) suggesting that this level might be a virtual molecular orbital due to the plasmon-induced coupling (fast electron transfer) between the two redox states of the Fc. Finally, in all our experiments with light a thermal effect can be excluded considering the low light intensity ($\approx$ 28 W/m$^2$) compared to previous experiments reporting bolometric plasmon enhanced photocurrent ($\approx$ 10$^7$ W/m$^2$).[12] Further theoretical works will be appealing to comfort the proposed plasmon-assisted mechanism for the suppression of the hysteresis and NDC behaviors.

## CONCLUSIONS.

We have observed the dynamic switching of the ferrocene moiety ($Fc^0$/$Fc^+$ states) in a double barrier molecular junctions when the ferrocene is connected to the Au electrode and the Ag nanocube through two insulating ultra thin layers (a self-assembled monolayer of alkylthiols and a PVP layer, respectively). This switching manifests in the current-voltage curves as a large hysteresis behavior at positive voltages and as a negative differential conductance behavior at negative voltages. We have suggested a phenomenological model to explain these behaviors.

Further experiments are now required to confirm this model, *e.g.*, dependence on the alkyl chain length, PVP thickness (thinner the barrier, faster the lifetime), effect of temperature (lower the temperature, slower the lifetime, if the switching is thermally activated), and voltage pulse experiments to determine the switching rates. This dual behavior open perspectives to use these Au-ferrocenyl-alkylthiol/AgNC molecular junctions as artificial synaptic devices for neuromorphic computing.[70] When light at plasmonic resonance is shined on the devices, the current hysteresis loop and the NDC behaviors are suppressed, likely due to plasmon-induced fast electron transfer between the two redox states (lifetime reduction of the states).

# Associated contents

**Supporting Information.**

The Supporting Information is available free of charge at https://pubs.acs.org/....

> Ellipsometry, topographic AFM images, protocol for conductive-AFM measurements, C-AFM contact area, number of molecules contacted, details on the fits of the I-V curves with the analytical SEL model; fits of the Fc switching and NDC behaviors, additional plots and control experiments, additional data under light illumination, XPS measurements and plasmonic simulations.

# Author information

**Corresponding author.**

**Dominique VUILLAUME** - *Institute for Electronics Microelectronics and Nanotechnology (IEMN), CNRS, Av. Poincaré, Villeneuve d'Ascq, France.* Orcid: 0000-0002-3362-1669, E-mail: dominique.vuillaume@iemn.fr

**Authors.**

Mariem BEN YOUSSEF - *Institute for Electronics Microelectronics and Nanotechnology (IEMN), CNRS, Av. Poincaré, Villeneuve d'Ascq, France.* Orcid: 0009-0002-5101-654X, E-mail: mariembenyoussef822@gmail.com

Peeranuch POUNGSRIPONG, - *Centre Interdisciplinaire de Nanoscience de Marseille (CINaM), CNRS, Aix Marseille Univ., Marseille, France.* Orcid: 0000-0002-3744-0929, E-mail: peeranuch.poungsripong@etu.univ-amu.fr

Hugo BIDOTTI - *Institut de Chimie Radicalaire (ICR), CNRS, Aix Marseille Univ., Marseille, France.* Orcid: 0000-0001-5547-9141, E-mail: bidottihugo@gmail.com

Halidou ABDOUL-YASSET - *Institut Matériaux Microélectronique Nanosciences de Provence (IM2NP), CNRS, Aix-Marseille Univ., Marseille, France.* Orcid: 0009-0000-0411-8135, E-mail: halidou.abdoul-yasset@im2np.fr

Didier GIGMES - *Institut de Chimie Radicalaire (ICR), CNRS, Aix Marseille Univ., Marseille, France.* Orcid: 0000-0002-8833-8393, E-mail: didier.gigmes@univ-amu.fr

Olivier MARGEAT - *Centre Interdisciplinaire de Nanoscience de Marseille (CINaM), CNRS, Aix Marseille Univ., Marseille, France.* Orcid: 0000-0003-3716-2399, E-mail: olivier.MARGEAT@univ-amu.fr

Beniamino SCIACCA - *Centre Interdisciplinaire de Nanoscience de Marseille (CINaM), CNRS, Aix Marseille Univ., Marseille, France.* Orcid: 0000-0003-1113-1391, E-mail: beniamino.SCIACCA@univ-amu.fr

Judikael Le ROUZO - *Institut Matériaux Microélectronique Nanosciences de Provence (IM2NP), CNRS, Aix-Marseille Univ., Marseille, France.* Orcid: 0000-0002-7907-3725, E-mail: judikael.le-rouzo@univ-amu.fr

David DUCHE - *Institut Matériaux Microélectronique Nanosciences de Provence (IM2NP), CNRS, Aix-Marseille Univ., Marseille, France.* Orcid: 0000-0001-5530-7876, E-mail: david.DUCHE@univ-amu.fr

***Author Contributions***

M.B.Y did all the electrical measurements. P.P., O.M. and B.S. synthesized the silver nanocubes and developed the deposition protocol on SAMs. H.B. and D.G. synthesized the molecules and the SAMs. H.A.Y. and J.L.R. prepared the $^{TS}$Au substrates. H.A.Y, J.L.R. and D.D. did the plasmonic simulations. M.B.Y. and D.V. analyzed the data. The manuscript was written by D.V. with the contributions and comments of all the authors. All authors have given approval of the final version of the manuscript.

***Note***

The authors declare no competing financial interest.

**Acknowledgements.**

We acknowledge the financial support from the ANR (project "Plasmore-light", # ANR-21-CE50-0037). We thank Mathieu Koudia (IM2NP) for the XPS measurements and the staff of the IEMM scanning probe microscopy platform for technical support.

**For Table of Contents Only**

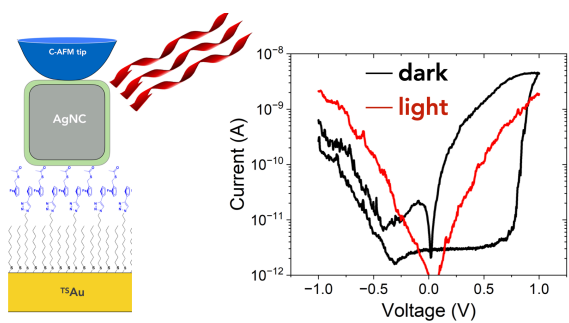

# Dynamic Switching of Ferrocene and Plasmonic Interactions in Au/Self-Assembled Monolayer/Single Ag Nanocube Molecular Junctions.

Mariem BEN YOUSSEF,[1] Peeranuch POUNGSRIPONG,[2] Hugo BIDOTTI,[3] Halidou ABDOUL-YASSET,[4] Didier GIGMES,[3] Olivier MARGEAT,[2] Beniamino SCIACCA,[2] Judikael Le ROUZO,[4] David DUCHE[4] and Dominique VUILLAUME.[1*]

1) Institute for Electronics Microelectronics and Nanotechnology (IEMN), CNRS, Av. Poincaré, Villeneuve d'Ascq, France.

2) Centre Interdisciplinaire de Nanoscience de Marseille (CINaM), CNRS, Aix-Marseille University, Marseille, France.

3) Aix-Marseille University, CNRS, Institut de Chimie Radicalaire (ICR), Marseille, France.

4) Institut Matériaux Microélectronique Nanosciences de Provence (IM2NP), CNRS, Aix-Marseille University, Marseille, France.

* Corresponding authors: dominique.vuillaume@iemn.fr

# Supporting Information.

## I. Ellipsometry.

We recorded spectroscopic ellipsometry data (on ca. 1 cm² samples) in the visible range using a UVISEL (Horiba Jobin Yvon) spectroscopic ellipsometer equipped with DeltaPsi 2 data analysis software. The system acquired a spectrum ranging from 2 to 4.5 eV (corresponding to 300–750 nm) with intervals of 0.1 eV (or 15 nm). The data were taken at an angle of incidence of 70°, and the compensator was set at 45°. We fit the data by a regression analysis to a film-on-substrate model as described by their thickness and their complex refractive indexes. First, a background for the substrate before monolayer deposition was recorded. We acquired three reference spectra at three different places of the surface spaced of few mm. Secondly, after the monolayer deposition, we acquired once again three spectra at three different places of the surface and we used a 2-layer model (substrate/SAM) to fit the measured data and to determine the SAM thickness. We employed the previously measured optical properties of the substrate (background), and we fixed the refractive index of the monolayer at 1.50.[1] We note that a change from 1.50 to 1.55 would result in less than a 1 Å error for a thickness less than 30 Å. The three spectra measured on the sample were fitted separately using each of the three reference spectra, giving nine values for the SAM thickness. We calculated the mean value from this nine thickness values and the thickness incertitude corresponding to the standard deviation. Overall, we estimated the accuracy of the SAM thickness measurements at ± 2 Å.[2]

## II. Topographic AFM images.

The SAMs were examined by topographic AFM (Fig. S1) using a tip probe (model LprobeTapping20 from Vmicro). The SAM surfaces are homogeneous and flat, they are free of gross defects (neither pinhole nor aggregate). For the $^{TS}$Au-S-C$_{8/11}$ SAM, the terraces are reminiscent of the underlying $^{TS}$Au substrate.[3] The white spots are dusts. Both are masked for the roughness analysis. The observed rms roughness value for the surface of the $^{TS}$Au-S-C$_{8/11}$ SAM is 0.39 nm, while the value of the $^{TS}$Au-S-C$_{8/11}$-Fc SAM is 0.66 nm (analyzed with Gwyddion[4]). Both values are close to the roughness observed for our $^{TS}$Au substrates (≈ 0.3-0.4 nm)[3,5] that indicates the good formation of a compact monolayer.

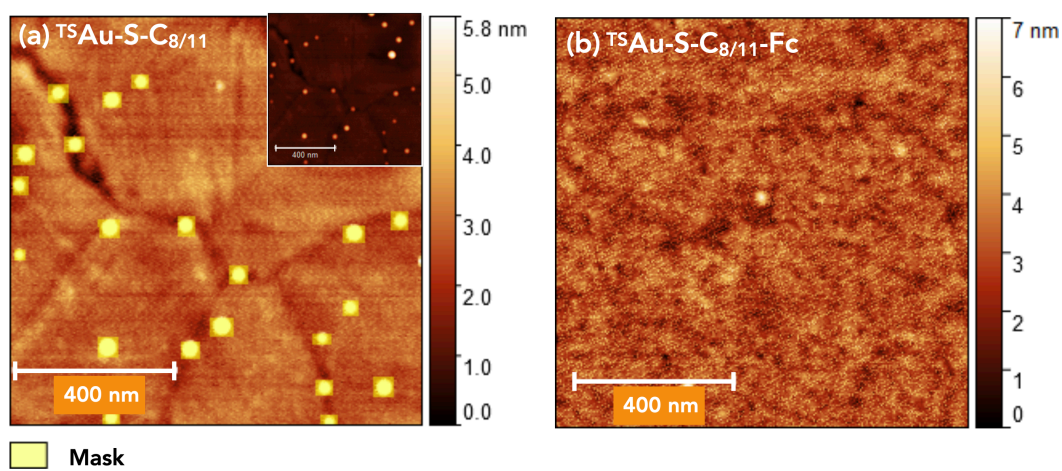

**Figure S1**. (a) topographic AFM image of the $^{TS}$Au-S-C$_{8/11}$ SAM. The inset is the unmasked image (b) topographic AFM image of the $^S$Au-S-C$_{8/11}$-Fc SAM.

### III. Conductive-AFM.

#### 1. Raw data analysis.

The raw set of I-V data was inspected and some I-V curves were discarded from the analysis: typically the I-V traces that reached the sensitivity limit (almost flat I-V traces and noisy I-Vs) and displayed random staircase behavior (due to the sensitivity limit of the equipment, typically few pA depending on the gain of the trans-impedance amplifier and the resolution of the analog-digital converter. At high currents, the I-V traces that reached the compliance level of the trans-impedance amplifier (depending on the gain of the amplifier) and/or I-V traces displaying large and abrupt steps during the scan (contact instabilities). The final number of I-V traces retained for analysis is indicated in the figure captions.

#### 2. C-AFM contact area, number of molecules contacted.

Considering: (i) the area per molecule on the surface and (ii) the estimated C-AFM tip contact surface (see below), we estimated the C-AFM tip contact area and the number, N, of molecules contacted by the tip as follows. As usually reported in literature[6-9] the contact radius, a, between the C-AFM tip and the SAM surface, and the SAM elastic deformation, δ, are estimated from a Hertzian model:[10]

$$a^2 = \left(\frac{3RF}{4E^*}\right)^{2/3} \tag{S1}$$

$$\delta = \left(\frac{9}{16R}\right)^{1/3}\left(\frac{F}{E^*}\right)^{2/3} \tag{S2}$$

with F the tip load force, R the tip radius (20 nm) and E* the reduced effective Young modulus defined as:

$$E^* = \left(\frac{1}{E^*_{SAM}} + \frac{1}{E^*_{tip}}\right)^{-1} = \left(\frac{1-v^2_{SAM}}{E_{SAM}} + \frac{1-v^2_{tip}}{E_{tip}}\right)^{-1} \tag{S3}$$

In this equation, $E_{SAM/tip}$ and $v_{SAM/tip}$ are the Young modulus and the Poisson ratio of the SAM and C-AFM tip, respectively. For the Pt/Ir (90%/10%) tip, we have $E_{tip}$ = 204 GPa and $v_{tip}$ = 0.37 using a rule of mixture with the known material data.[11] These parameters for alkyl-ferrocene SAMs are not known and, in general, they are not easily determined in such a monolayer material. Thus, we consider the value of an effective Young modulus of the SAM $E^*_{SAM}$ = 38 GPa as determined for the "model system" alkylthiol SAMs from a combined mechanic and electron transport study.[8] With these parameters and F ≈ 3-5 nN, we estimated $a$ ≈ 1.2 nm (contact area ≈ 4.7 nm²) and $\delta$ ≈ 0.07 nm (with a mean F ≈ 4 nN). From previously reported CV measurements on the same $^{TS}$Au-S-$C_{8/11}$-Fc SAMs made with the same "click" chemistry approach,[12] the density of ferrocene moieties was 1.7x10$^{14}$ Fc/cm², i.e. an area per Fc molecule of ca. 0.6 nm². By calculating the maximum packing of a 4.7 nm² circle by smaller circles of 0.6 nm² (see http://packomania.com/), we infer that about 5 molecules are measured in the SAM/PtIr junction for the $^{TS}$Au-S-$C_{8/11}$-Fc SAMs. For the $^{TS}$Au-S-$C_{8/11}$ SAMs, we used the known area per alkylthiol of ≈ 0.3 nm² (at close packing)[1, 13] that gives N ≈ 10. However, we note that the values of the Young modulus of the SAMs reported in the literature are spread. For instance a value of 1.4 GPa was also reported for alkylthiol SAMs.[14] With this value, the estimated contact area increases to 38 nm² (all the other parameters being identical) and the N increases to ≈ 50 and ≈ 100 (ferrocene and alkykthiol, respectively).

For the SAMs contacted by the AgNC (nominal side of 60 nm), we roughly estimate that the "contact" area is the area of the NC face, assuming a flat face entirely in contact with the SAM. Under this approximation, we get N ≈ 5100 for the $^{TS}$Au-S-$C_{8/11}$-Fc SAMs and N ≈ 10$^4$ for the $^{TS}$Au-S-$C_{8/11}$ SAMs.

We note that the exact value of N does not have a large influence on the fitted parameter $\varepsilon_0$ that is mainly fixed by the voltage at which the current deviates from the linear ohmic behavior (at low voltages) to the super-linear shape of arctan functions. The value of N has a large influence on the values of the molecule/electrode coupling energy $\Gamma_1$ and $\Gamma_2$ values (Eq. S4), which, due to the crude determination of N, must be considered are "effective" coupling energies of the SAM with the electrodes and they are used only for a relative comparison of the SAMs measured with the same C-AFM conditions in the present work and they cannot be used for a direct comparison with other reported data (as for example from single molecule experiments).

### 3. Fits of the I-V curves with the analytical SEL model.

All the I-V traces of the data set of the molecular junctions were fitted with the single-energy level (SEL) model, which supposes that a single molecular orbital dominates the charge transport, the voltage mainly drops at the molecule/electrode interface and that the molecular orbital broadening due to the coupling with the electrodes is described by a Lorentzian or Breit-Wigner distribution. The model is described by the following analytical expression:[15, 16]

$$I(V) = N \frac{8e}{h} \frac{\Gamma_1 \Gamma_2}{\Gamma_1 + \Gamma_2} \left[ arctan\left(\frac{\epsilon_0 + \frac{\Gamma_1}{\Gamma_1+\Gamma_2} eV}{\Gamma_1 + \Gamma_2}\right) - arctan\left(\frac{\epsilon_0 - \frac{\Gamma_2}{\Gamma_1+\Gamma_2} eV}{\Gamma_1 + \Gamma_2}\right) \right] \quad (S4)$$

with $\varepsilon_0$ the energy of the molecular orbital involved in the transport (with respect to the Fermi energy of the electrodes), $\Gamma_1$ and $\Gamma_2$ the electronic coupling energy between the molecular orbital and the electron clouds in the two electrodes, e the elementary electron charge, h the Planck constant and N the number of molecules contributing to the electron transport in the molecular junction (assuming independent molecules conducting in parallel, *i.e.*, no intermolecular interaction[17-19]). N is estimated to ≈ 5100 when the $^{TS}$Au-S-C$_{8/11}$-Fc SAMs are connected by the AgNC and ≈ 5 when the C-AFM tip is directly contacting the SAM (*vide supra*), N ≈ 10$^4$ and ≈ 10 for the $^{TS}$Au-S-C$_{8/11}$ SAMs, respectively.

The SEL model is strictly valid at 0 K, since the Fermi-Dirac electron distribution of the electrodes is not taken into account. However, it was shown that it can be reasonably used to fit data measured at room temperature for voltages below the transition between the off-resonant and resonant transport conditions at which the broadening of the Fermi function modify the I-V shape leading to sharpened increase of the current.[20-22] We defined this bias voltage window of confidence by using the transition voltage spectroscopy (TVS)[23-27] that give us an estimate of this transition regime. Plotting $|V^2/I|$ vs. V (Fig. S2a),[28] we determined the transition voltages $V_{T+}$ and $V_{T-}$ for both voltage polarities at which the bell-shaped curve is maximum (or we assume $V_{T+} = V_{T-}$ when only one polarity of the I-Vs can be used). The value of $\varepsilon_{0\text{-TVS}}$ is estimated by:[27]

$$|\varepsilon_{0-TVS}| = 2 \frac{e|V_{T+}V_{T-}|}{\sqrt{V_{T+}^2 + 10|V_{T+}V_{T-}|/3 + V_{T-}^2}}$$

(S5).

Figure S2 illustrates the procedure (for the $^{TS}$Au-S-C$_{8/11}$-Fc/C-AFM tip junctions). From the TVS plot, the maxima of the almost bell-shaped curves were determined by fitting a 2$^{nd}$ order polynomial function around the max of the bell-shaped curves (to cope with noise and distorted curves due to the dispersion of the data sets). These maxima (red dashed lines) give the transition voltages $V_{T-}$ = -0.48 V and $V_{T+}$ = 0.59 V, from which we estimate the energy position of the MO using Eq. S5. Then, we fixed the bias window of confidence by solving a numerical analysis reported in Ref. 29 (Eq. 17 in Ref. 29) that gives the condition of applicability of the 0K SEL model to room temperature experimental data. In the present case, we obtain $|V| < 0.65$ V. Then all the I-V traces of this data set were fit within the bias window from -0.65 to 0.65 V. Figure S2 gives a typical fit of the SEL model on the mean Ī-V of the $^{TS}$Au-S-C$_{8/11}$-Fc/C-AFM tip junctions. We note that, with these conditions, the two methods give values for the energy level $\varepsilon_0$ in reasonable agreement. The fits of the SEL model were done with the routine included in ORIGIN software (version 2023, OriginLab Corporation, Northampton, MA, USA), using the method of least squares and the Levenberg Marquardt iteration algorithm. In several cases, the TVS plots do not give maxima, *e.g.* Fig. S3. In these cases, the SEL model was fit on the full-voltage range (-1 to 1 V).

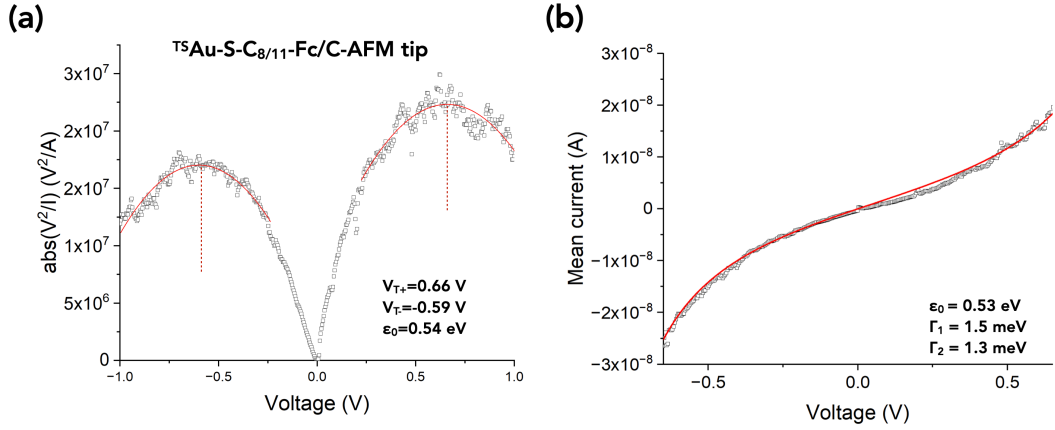

*Figure S2. (a)* Data of the mean Ī-V of the $^{TS}$Au-S-C$_{8/11}$-Fc/C-AFM tip junction plotted as $|V^2/I|$ vs. V. The red lines are the fit by a 2$^{nd}$ order polynomial function used to determine the voltage positions of the maxima. *(b)* Fit (blue curve) of the SEL model (Eq. S4) on the mean Ī-V of $^{TS}$Au-S-C$_{8/11}$-Fc/C-AFM tip junction, fit limited between -0.65 and 0.65 V (see text), the fit parameters are given in the panel.

### *4. Fits of the Fc switching and NDC behaviors.*

We used the analytical model developed in Ref. 30. For convenience, we rewrite below the equations to cope with the direction of the applied voltage in our experiments with respect to the convention used in Ref. 30. For the part II of the I-V (Fc$^+$ → Fc$^0$ switching) at V > 0, we used Eq. 12 in Ref. 30 to calculate the current I, which is rewritten as

$$I = A \frac{R_{Ag,+/0} R_{Au,0/+}}{R_{Ag,+/0} + R_{Au,0/+}} \quad (S6)$$

with $R_{Ag,+/0}$ and $R_{Au,0/+}$ are the electron transport transfer rate from the Ag NC electrode to switch from the Fc$^+$ to Fc$^0$ states (injection into the molecule), and the electron transport rate to the Au electrode to switch from the Fc$^0$ to Fc$^+$ states (electron removal from the molecule), respectively. We used the simplified analytical equations (33, 34) in Ref. 30, rewritten as

$$R_{Ag,+/0} \cong \frac{\gamma_{Ag}}{2} erfc\left(\frac{\lambda + \varepsilon - 0.5V}{2\sqrt{\lambda kT}}\right) \quad (S7)$$

$$R_{Au,0/+} \cong \frac{\gamma_{Au}}{2} erfc\left(\frac{\lambda - \varepsilon - 0.5V}{2\sqrt{\lambda kT}}\right) \quad (S8)$$

In these equations, $\gamma_{Ag}$ and $\gamma_{Au}$ are the molecule-metal coupling energies, ε the energy difference between the two states (with respect to the Fermi energy), λ the reorganization energy, k the Boltzmann constant, T the temperature. To cope with the real size of the MJ (not considered in Ref. 30), we added a parameter A in Eq. (S6) that is a scaling factor to adjust the level of the current, which depends on the size of electrodes and the exact number of molecules in the MJ. Given the large number of parameters, we fixed $\gamma_{Ag} = \gamma_{Au} = 1$ meV (weak electrode coupling).

For the NDC, we need to use the two conductance channel model.[30, 31] We used the analytical equation (28) in Ref. 30. Since the experimental NDC is observed at V < 0, while the simulations in Ref. 30 only consider positive bias, the two electrodes have been inverted. The equation reads

$$I_{NDC} = \frac{I}{1 + \frac{R_{Au,0/+}}{R_{Ag,+/0}+R_{Au,0/+}} exp\left(\frac{-\varepsilon_2+0.5V}{kT}\right)} \quad (S9)$$

where I is given by Eq. (S6) and $\varepsilon_2$ is the energy level of the second conduction channel.

## V. Additional plots.

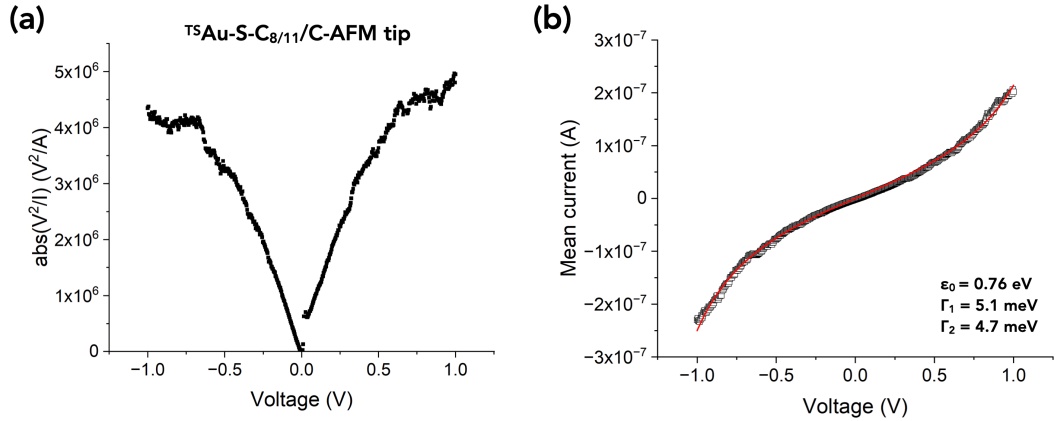

***Figure S3. (a)*** *Data of the mean Ī-V of the $^{TS}$Au-S-C$_{8/11}$/C-AFM tip junction plotted as |V²/I| vs. V. No maxima are clearly detected. **(b)** Fit (red curve) of the SEL model (Eq. S4) on the mean Ī-V of $^{TS}$Au-S-C$_{8/11}$/C-AFM tip junction, the fit parameters are given in the panel.*

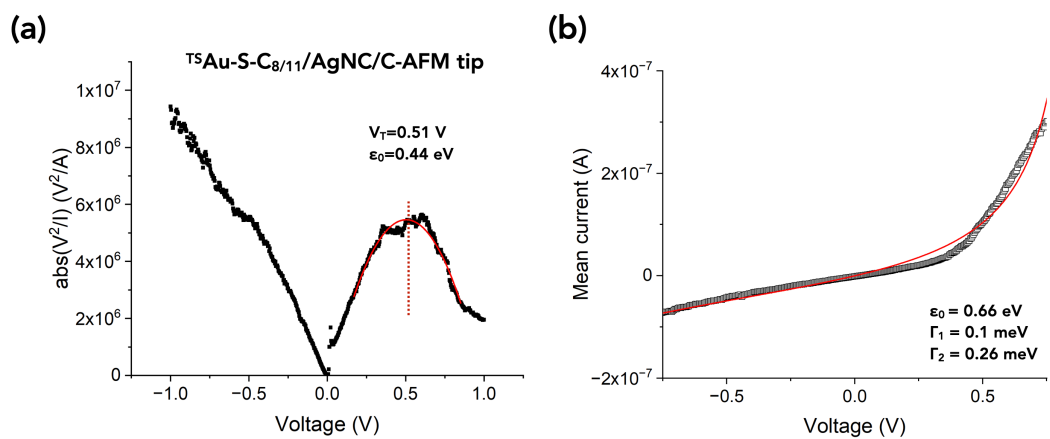

***Figure S4. (a)*** *Data of the mean Ī-V of the* $^{TS}$*Au-S-C$_{8/11}$/AgNC/C-AFM tip junction plotted as* $|V^2/I|$ *vs. V. Only one maximum is detected.* ***(b)*** *Fit (red curve) of the SEL model (Eq. S4) on the mean Ī-V of* $^{TS}$*Au-S-C$_{8/11}$/AgNC/C-AFM tip junction fit limited between -0.75 and 0.75 V (vide supra), the fit parameters are given in the panel.*

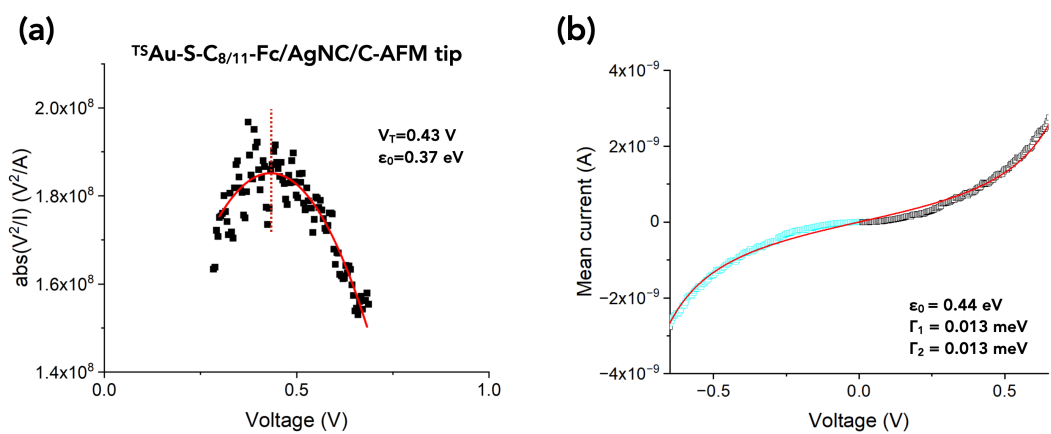

***Figure S5. (a)*** *Data of the mean Ī-V of the* $^{TS}$*Au-S-C$_{8/11}$-Fc/AgNC/C-AFM tip junction plotted as* $|V^2/I|$ *vs. V. Only the backward trace from 1 to 0 V is used (see main text).* ***(b)*** *Fit (red curve) of the SEL model (Eq. S4) on the mean Ī-V of* $^{TS}$*Au-S-C$_{8/11}$/AgNC/C-AFM tip junction. Since an accurate fit of Eq. 1 requires to have data at positive and negative voltages (for the two arctan functions) a symmetric negative branch was added (light blue) by inverting the measured positive branch. The agreement is good with the TVS determination. The fit parameters are given in the panel.*

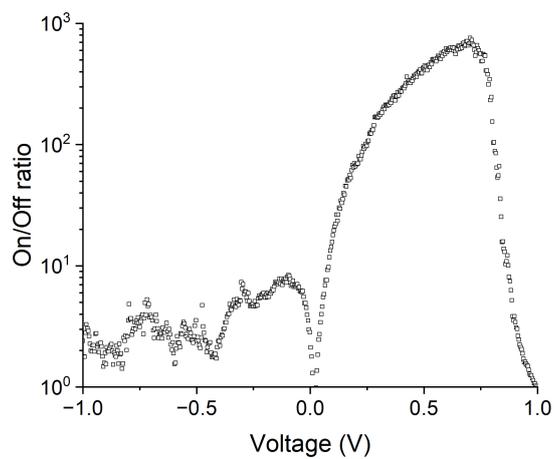

*Figure S6.* Mean on/off ratio ($R_{ON/OFF}$) calculated from the mean $\bar{I}$-V traces of Fig. 2a.

## VI. Control experiments

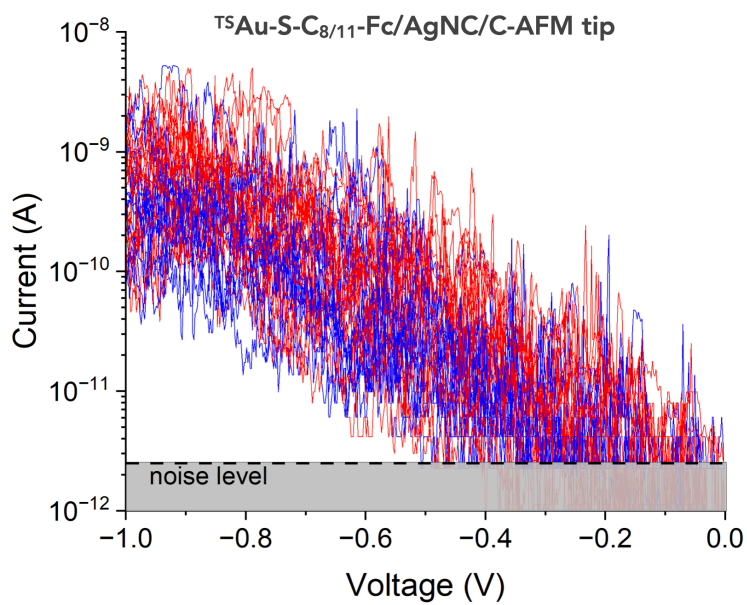

*Figure S7.* I-V data set (35 traces) of the $^{TS}$Au-S-C$_{8/11}$-Fc/AgNC/C-AFM tip junction back and forth from - 1 V to 0 V.

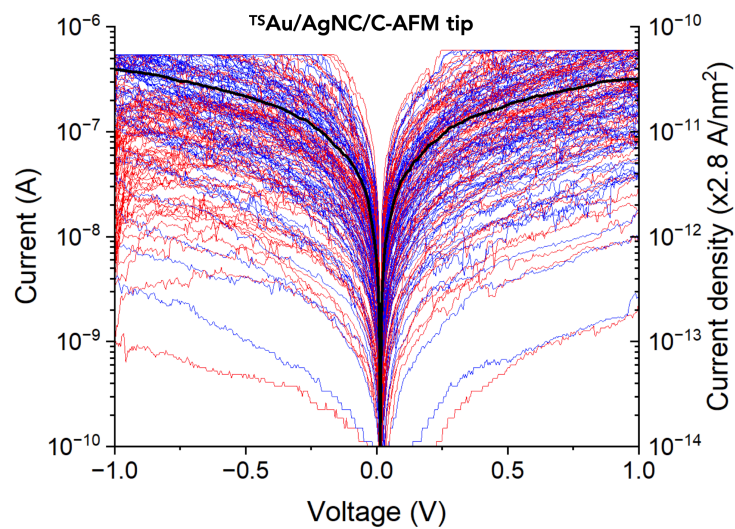

**Figure S8.** I-V data set (142 traces) of the $^{TS}$Au/AgNC/C-AFM tip junction, forward -1 to 1 V (red) and backward 1 to -1 V (blue) traces. The black line is the mean $\bar{I}$-V from all the traces. Note that the mean $\bar{I}$-V is underestimated since many traces saturates (compliance of the trans-impedance preamplifier). There is no hysteresis, nor NDC, for these Au/PVP/Ag tunnel junctions.

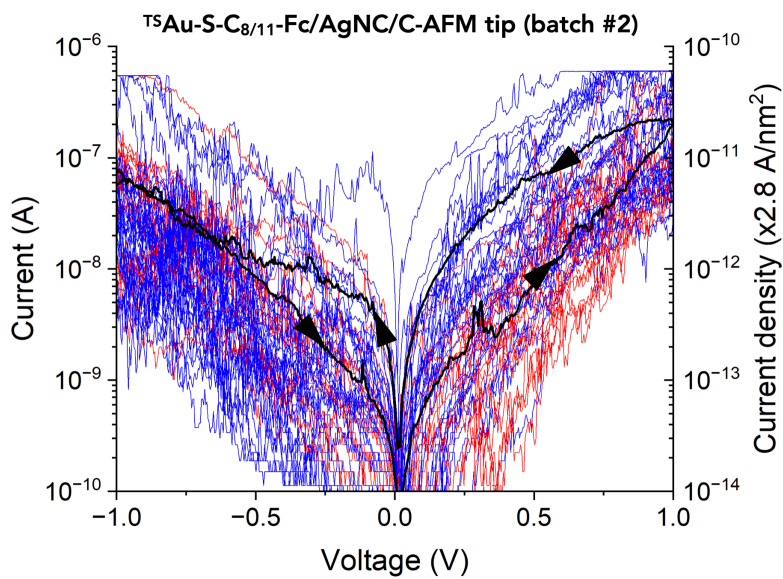

**Figure S9.** I-V data set (53 traces) of the $^{TS}$Au-S-C$_{8/11}$-Fc/AgNC/C-AFM tip junctions (batch #2). The dark lines are the mean $\bar{I}$-V for the forward (red) and backward (blue) traces.

# VII. Additional data under light illumination.

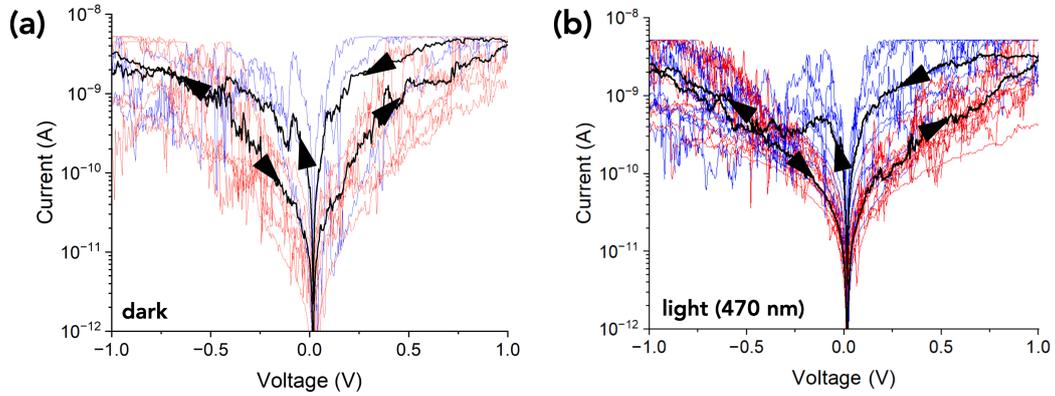

***Figure S10. (a)*** *Typical I-V curves in the dark of the $^{TS}$Au-S-C$_{8/11}$-Fc/AgNC/C-AFM tip junctions (same sample but other NCs than in Fig. 5). Only a few traces (10) were recorded to check the hysteresis and NDC. The dark lines are the mean $\bar{I}$-V for the forward (red) and backward (blue) traces.* ***(b)*** *I-V data set (21 traces) of the $^{TS}$Au-S-C$_{8/11}$-Fc/AgNC/C-AFM tip junction under light illumination at 470 nm (bandwidth : FWHM 20 nm). The dark lines are the mean $\bar{I}$-V for the forward (red) and backward (blue) traces.*

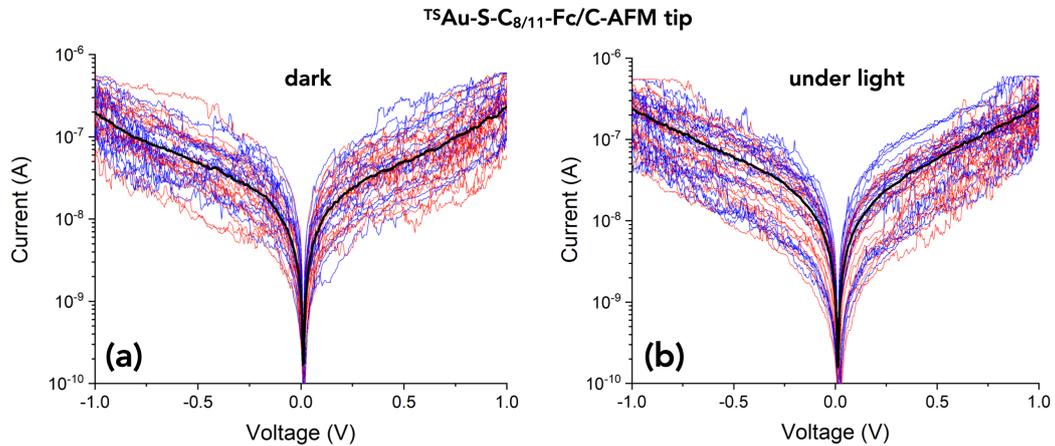

***Figure S11. (a)*** *I-V data set (40 traces) in the dark of the $^{TS}$Au-S-C$_{8/11}$-Fc/C-AFM tip junctions. The dark line is the mean $\bar{I}$-V for all the traces.* ***(b)*** *I-V data set (40 traces) of the $^{TS}$Au-S-C$_{8/11}$-Fc/C-AFM tip junction under light illumination (470-850 nm wavelength). The back line is the mean $\bar{I}$-V from all the traces.*

## VIII. XPS measurements.

XPS measurements (Omicron XM1000 source fitted in an UHV chamber with a residual pressure of $10^{-10}$ mbar) were recorded with a monochromatic Al$_{K\alpha}$ X-ray source (hν = 1486.6 eV), a detection angle of 20° as referenced to the sample surface (Omicron EA125 analyzer) and an analyzer pass energy of 30 eV. Background was subtracted by the Shirley method.[32] The peaks were decomposed using Voigt functions and a least squares minimization procedure. Binding energies were referenced to the C 1s BE, set at 284.8 eV.

The S-Au bonding is confirmed by the S2p region showing the expected two doublets (S2p$_{1/2}$ and S2p$_{3/2}$) associated to the sulfur bonded to gold (S2p$_{1/2}$ at 162.3 eV, S2p$_{3/2}$ at 161.1 eV) and the S atoms unbounded to gold (S2p$_{1/2}$ at 163.2 eV, S2p$_{3/2}$ at 161.9 eV) (Fig. S12a). These doublets are separated by ≈1.2 eV as expected with an amplitude ratio [S2p$_{1/2}$]/[S2p$_{3/2}$] set at 1/2.

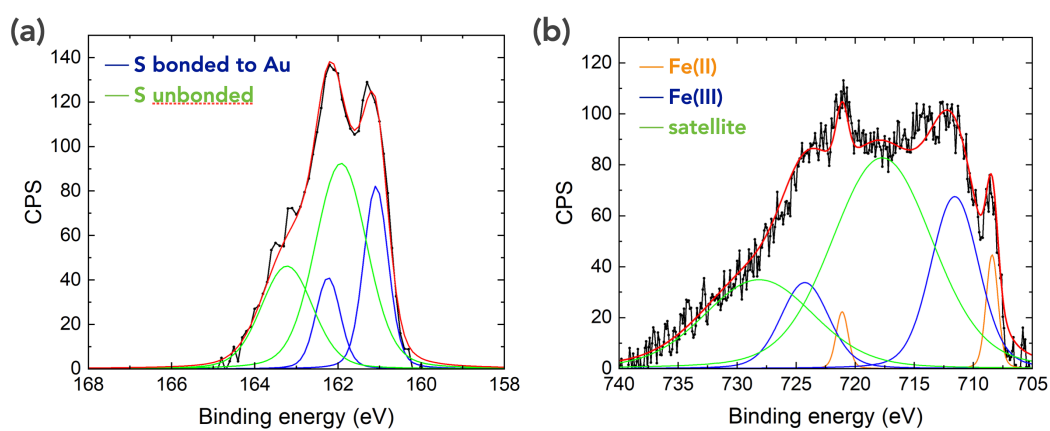

*Figure S12*. *(a) XPS spectra of the S2p region. The dark line is the measurement. The red line is the envelope fit with the two doublets decomposition shown in blue (S bonded to Au) and green (unbonded S) - see text for details.* ***(b)*** *Fe2p region. The dark line is the measurement. The red line is the envelope fit taking into account the contributions (doublets) from the Fe(II) and Fe(III) oxidation states, in orange and blue, respectively, and shakeup satellite peaks (see text for details)*

In the Fe2p region (Fig. S12b), we observed a complex signal corresponding to a mix of the Fe(II) and Fe(III) oxidation states. Each chemical specie is fit with the 2p$_{1/2}$ and 2p$_{3/2}$ doublet constrained to have a 1:2 peak area ratio and a peak separation of 13.1 eV. The doublet at 708.4 eV and 721.5 eV (orange line, Fig. S12b) is associated to Fe(II)2p$_{3/2}$ and Fe(II)2p$_{1/2}$, respectively.

The doublet associated to Fe(III)$2p_{3/2}$ and Fe(III)$2p_{1/2}$ is observed at 711.6 eV and 724.7 eV (blue line in Fig. S12b), respectively. The envelope fit is completed by two satellite peaks at 717.7 (Fe $2p_{3/2}$) and 728.1 eV (Fe $2p_{1/2}$).[33-35] From the integrated peak areas of the Fe(III) and Fe(II) doublets, we estimated a ratio [Fe(III)]/[Fe(II)] ≈ 6.

## IX. Plasmonic simulations.

The mapping of the vertical axis electric field was generated using the Finite-Difference Time-Domain method with the commercial software Lumerical (www.lumerical.com). The material properties for gold and silver were taken from experimental data by Johnson and Christy.[36] The dielectric constant of PVP was set to 1.52, and the refractive index of the SAM was set to 1.46.[12, 37] Our boundary conditions were set to be periodic with a period of 200 nm, large enough to avoid interactions between nanocubes, but small enough to prevent grating effects within the resonance range.[38] A Perfectly Matched Layer was applied along the z-axis to minimize reflections. The horizontal meshing was set to 0.2 nm to accurately model the thin PVP coating layer. The incident electric field was applied along the x-axis, oriented normally to the surface. The electric field mapping was conducted at resonance. Given the measured thicknesses of the Fc SAMs and PVP layers (2.8 ± 0.2 nm, *vide supra* and 1.5±0.9 nm)[39], we did the simulations for two extrema of the thickness of the cavity (SAM+PVP), a minimum at 3.2 nm and a maximum at 5.4 nm. The wavelengths of the fundamental gap plasmon modes were determined to be 665 nm for the thicker structure and 755 nm for the thinner structure as shown in the reflection spectra in Fig. S13b. The mapping is shown at these wavelengths for both structures at the Ag/PVP interface (black dashed line in Fig. S13a) and at the PVP/SAM interface (red dashed line in Fig. S13a). The electric-field amplification factor ranges between ≈ 70 and ≈ 140 (Fig. S14).

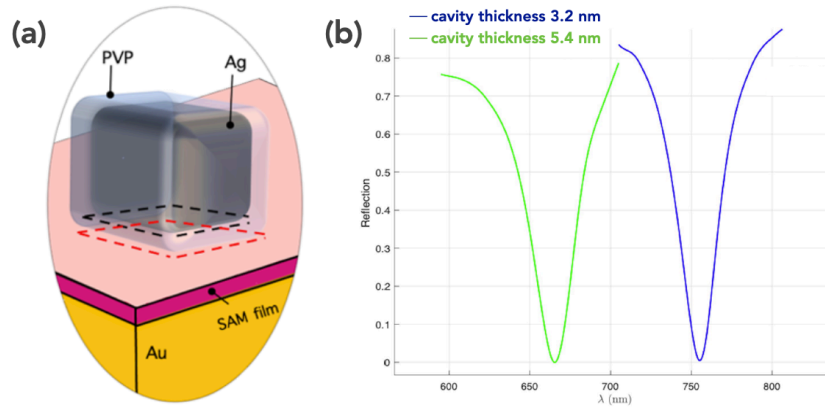

*Figure S13. (a)* Scheme of the simulated device. The dashed lines indicates the plane for the mapping of the electric-filed enhancement shown in Fig. S14: black (red) dashed lines at the Ag/ PVP (PVP/SAM) interfaces, respectively. *(b)* Reflection spectra for the two cavity thicknesses (see text) given the fundamental plasmon mode.

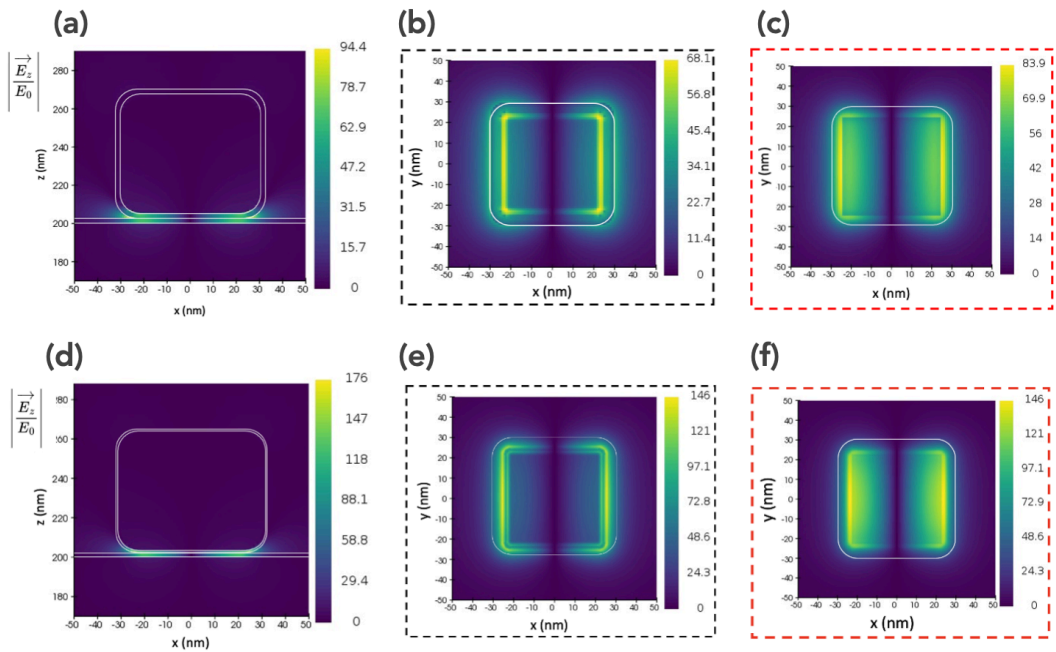

*Figure S14.* Mapping of the electric field amplification factor for the thickest cavity *(a-c)* at 665 nm and the thinnest one *(d-f)* at 755 nm. Panels *(a)* and *(b)* are side views. Panels *(b)* and *(e)* are the *top views showing the mapping at the Ag/PVP interface (dashed black line in Fig. S13a), panels **(c)** and **(f)** at the PVP/SAM interface (dashed red line in Fig. S13a).*

## References.